\documentclass[a4paper,10pt,preprintnumbers,floatfix,twocolumn,superscriptaddress,pra,bibnotes,showpacs,notitlepage,longbibliography,aps]{revtex4-2}
\usepackage[utf8]{inputenc}
\usepackage[T1]{fontenc}
\usepackage[colorlinks,breaklinks,linkcolor=blue,citecolor=magenta,urlcolor=magenta]{hyperref}
\usepackage{amsmath,amsthm,amssymb,amsfonts,dsfont,mathtools}
\usepackage{graphicx}
\usepackage{physics}
\usepackage{xcolor}
\usepackage{float}

\renewcommand{\vec}[1]{\boldsymbol{#1}}
\newcommand{\id}{\ensuremath{\mathds{1}}}
\renewcommand{\d}{\mathsf{d}}

\begin{document}
\title{Bounded-Error Quantum Simulation via Hamiltonian and Lindbladian Learning}

\author{Tristan~Kraft}
\affiliation{Technical University of Munich, TUM School of Natural Sciences, Physics Department, 85748 Garching, Germany}
\affiliation{Munich Center for Quantum Science and Technology (MCQST), Schellingstrasse 4, 80799 Munich, Germany}
\affiliation{Institute for Theoretical Physics, University of Innsbruck,  Innsbruck, Austria}

\author{Manoj~K.~Joshi}
\affiliation{Institute for Quantum Optics and Quantum Information, Austrian Academy of Sciences, Innsbruck, Austria}
\affiliation{Institute for Experimental Physics, University of Innsbruck, Innsbruck, Austria}

\author{William~T.~Lam}
\affiliation{Univ.~Grenoble Alpes, CNRS, LPMMC,  Grenoble, France}

\author{Tobias~Olsacher}
\affiliation{Forschungszentrum Jülich GmbH, Peter Grünberg Institute, Quantum Control (PGI-8), 52425 Jülich, Germany}
\affiliation{Institute for Theoretical Physics, University of Innsbruck,  Innsbruck, Austria}
\affiliation{Institute for Quantum Optics and Quantum Information, Austrian Academy of Sciences, Innsbruck, Austria}

\author{Florian~Kranzl}
\affiliation{Institute for Quantum Optics and Quantum Information, Austrian Academy of Sciences, Innsbruck, Austria}
\affiliation{Institute for Experimental Physics, University of Innsbruck,  Innsbruck, Austria}

\author{Johannes~Franke}
\affiliation{Institute for Quantum Optics and Quantum Information, Austrian Academy of Sciences, Innsbruck, Austria}
\affiliation{Institute for Experimental Physics, University of Innsbruck,  Innsbruck, Austria}

\author{Lata~Kh~Joshi}
\affiliation{SISSA -- International School for Advanced Studies, Trieste, Italy}

\author{Rainer~Blatt}
\affiliation{Institute for Quantum Optics and Quantum Information, Austrian Academy of Sciences, Innsbruck, Austria}
\affiliation{Institute for Experimental Physics, University of Innsbruck, Innsbruck, Austria}

\author{Augusto~Smerzi}
\affiliation{
INO-CNR and LENS, Largo Enrico Fermi 2, 50125 Firenze, Italy}

\author{Daniel~Stilck~Fran\c{c}a}
\affiliation{Department of Mathematical Sciences and Quantum for Life Center, University of Copenhagen,  Copenhagen, Denmark}

\author{Beno\^\i t~Vermersch}
\affiliation{Univ.~Grenoble Alpes, CNRS, LPMMC,  Grenoble, France}
\affiliation{Quobly, Grenoble, France}

\author{Barbara~Kraus}
\affiliation{Technical University of Munich, TUM School of Natural Sciences, Physics Department, 85748 Garching, Germany}
\affiliation{Munich Center for Quantum Science and Technology (MCQST), Schellingstrasse 4, 80799 Munich, Germany}

\author{Christian~F.~Roos}
\thanks{These authors jointly supervised this work.}
\affiliation{Institute for Quantum Optics and Quantum Information, Austrian Academy of Sciences, Innsbruck, Austria}
\affiliation{Institute for Experimental Physics, University of Innsbruck, Innsbruck, Austria}

\author{Peter~Zoller}
\thanks{These authors jointly supervised this work.}
\affiliation{Institute for Theoretical Physics, University of Innsbruck,  Innsbruck, Austria}
\affiliation{Institute for Quantum Optics and Quantum Information, Austrian Academy of Sciences,  Innsbruck, Austria}

\begin{abstract}
Analog Quantum Simulators offer a route to exploring strongly correlated many-body dynamics beyond classical computation, but their predictive power remains limited by the absence of quantitative error estimation. Establishing rigorous uncertainty bounds is essential for elevating such devices from qualitative demonstrations to quantitative scientific tools. Here we introduce a general framework for bounded-error quantum simulation, which provides predictions for many-body observables with experimentally quantifiable  uncertainties. The approach combines Hamiltonian and Lindbladian Learning—a statistically rigorous inference of the coherent and dissipative generators governing the dynamics—with the propagation of their uncertainties into the simulated observables, yielding confidence bounds directly derived from experimental data. We demonstrate this framework on trapped-ion quantum simulators implementing long-range Ising interactions with up to 51 ions. We analyze error bounds on two levels. First, we learn an open-system model from experimental data collected in an initial time window of quench dynamics, simulate the corresponding master equation, and quantitatively verify consistency between theoretical predictions and measured dynamics at long times. Second, we discuss short-time error bounds directly from experimental measurements alone, without relying on classical simulation—crucial for entering regimes of quantum advantage. In both cases, the learned models reproduce the experimental evolution within the predicted bounds, demonstrating quantitative reliability and internal consistency. By integrating statistical learning, open-system modeling, and precise experimental control, bounded-error quantum simulation provides a scalable foundation for trusted analog quantum computation, bridging the gap between experimental quantum platforms and predictive many-body physics. The techniques presented here directly extend to digital quantum simulation.
\end{abstract}

\maketitle

\section{Introduction}

Quantum simulation holds the promise of efficiently solving complex
quantum many-body problems using programmable quantum devices, potentially
beyond what can be simulated classically~\cite{Gross2023,Daley2022}. The central goal is the
accurate prediction of expectation values of many-body observables
under both equilibrium and non-equilibrium dynamics. While substantial
progress has been achieved across atomic and superconducting platforms
operating in analog or digital modes~\cite{Blatt2012,Gross2017,Altman2021,Monroe2021,Fauseweh2024,Browaeys2020,Adler2024,Haghshenas2025,Manovitz2025,Phasecraft2025,Phasecraft22025,Mark2025,Abanin2025}, a critical challenge remains:
elevating quantum simulators into quantitative computational tools
that deliver not only estimates for observables, but also rigorous error bounds~\cite{Cai2024,Trivedi2024,Aharonov2025,Carrasco2021,Altman2021,Rao2025}. This challenge is particularly acute for analog
quantum simulators, which implement target Hamiltonians natively in
controllable many-body systems involving a large number of particles,
yet, as Noisy Intermediate Scale Quantum (NISQ) devices~\cite{Gross2023} lack the benefits of quantum error correction. Establishing a framework of bounded error (analog) quantum simulations, where predictions are accompanied
by rigorous error bounds which remain applicable in the regime of
quantum advantage~\cite{Carrasco2021,Daley2022,Trivedi2024,Kashyap2025}, is essential for advancing from proof-of-principle
demonstrations to quantitative scientific computation. 

Below, we describe a general framework for Bounded-Error Quantum Simulation (BEQS),
broadly applicable across diverse platforms, and demonstrate its experimental
implementation for real-time (quench) dynamics using a trapped-ion
analog quantum simulator. Our protocol proceeds in two steps: 

In the first step, we consider the quantum simulator as a “black box” in which the Hamiltonian acts as a latent variable that cannot be measured directly but needs to be inferred from experimental observations. Determining the parameters of this latent variable is generally known as an \emph{inverse statistical problem}, which is ubiquitous in science and engineering~\cite{Nguyen2017,Waqar2023}. In quantum simulation, this is realized through Hamiltonian learning~\cite{QiRanard2019,Bairey2019,Evans2019,Bentsen2019,Franca2022,Olsacher2025,Lam2026,Huang2023,Kokail2021,Joshi2023exploring,Guo2025,Franceschetto2025,Hu2025,Holzapfel2015,Hangleiter2024,Wilde2022}, in which the operator structure of the many-body Hamiltonian can be inferred efficiently from experimental data. As in Bayesian statistical inversion, this includes learning an error model that encapsulates our uncertainty about the model parameters~\cite{Bairey2019,Waqar2023}. For weakly dissipative systems governed by Markovian noise, the framework naturally extends to learning the Lindbladian of a master equation model~\cite{Bairey2020,Franca2022,Olsacher2025,Lam2026,Pastori2022}.

In the second step, the quantum simulator is operated as a computational device. The general quantum simulation task is to compute the expectation values of many-body observables, given a Hamiltonian as input, through measurements performed on the experimental system. In our framework, the relevant Hamiltonian (and Lindbladian) has been experimentally characterized, up to statistical uncertainties, by learning in the first stage. The quantum simulator then provides predictions for the observables under this learned model. BEQS seeks to establish rigorous error bounds on these predicted observables, directly linked to the statistical uncertainties in the learned Hamiltonian and Lindbladian.
Within these uncertainties, the simulator can provide predictions quantified by computable error bounds.
In regimes where classical simulations remain feasible, these error bounds follow naturally from propagating uncertainties in the learned model to the measured observables, enabling verification of the simulator’s performance. Importantly, as we show below, such error bounds~\cite{Cai2024} can also be inferred directly from experimental data—an essential capability when entering regimes beyond the reach of classical computation.

The BEQS protocol described above relies on a minimal set of control requirements standard to {\em programmable} quantum simulators~\cite{Blatt2012,Gross2017,Altman2021,Monroe2021,Fauseweh2024,Browaeys2020,Adler2024,Haghshenas2025,Manovitz2025,Mark2025,Abanin2025}. Aside from the reproducible implementation of a fixed target Hamiltonian $H$ without temporal drift, the only required capabilities are the preparation of arbitrary product states and high-fidelity product measurements in arbitrary bases. These assumptions are met across leading quantum-simulation platforms, and no further coherent multi-qubit control is required for state preparation or readout. We provide experimental demonstrations using a 1D trapped-ion quantum simulator with up to $51$ ions.

Quantum simulation has emerged as particularly promising applications for noisy intermediate-scale quantum (NISQ) devices, owing to their inherent robustness to errors \cite{GonzalezGarcia2022,Trivedi2024,Kashyap2025,Daley2022}.
BEQS contributes to the broader effort to render quantum simulations quantitatively reliable and verifiable. However, BEQS as defined above is different from existing verification and validation protocols (for recent reviews and tutorial see~\cite{ BlumeKohout2025,
Eisert2020, Gheorghiu2019, KlieschRoth2021} and, e.g., \cite{ Xiao2022, Jackson2024PNAS, JacksonDatta2025, Mills2025, Liu2025, Carrasco2021} and references therein). It does not determine an error between any ideal (noiseless) and the real  evolution, but rather determines the uncertainty about the outcome one has using the quantum simulator. That is, BEQS verifies quantum simulation in the sense that it determines error bounds on the expectation values of observables after a specified evolution time. Crucially the bounds, or uncertainty intervals, are determined directly from experimental data, without requiring any classical simulation—an essential feature for BEQS in regimes where quantum advantage is expected. In our analysis, we are mainly concerned with the statistical uncertainties arising from finite measurement resources. The protocols developed here are tailored to present-day NISQ devices, while remaining compatible with future fault-tolerant architectures.

\section{Motivation and Overview of Theory}

\begin{figure*}
    \centering
    \includegraphics[width=1\linewidth]{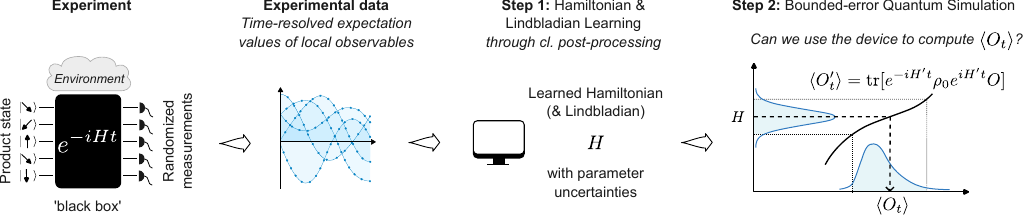}
    \caption{{\bf Overview of the BEQS protocol}: The analog quantum simulator is treated as a "black-box" device. Product input states are prepared, evolved for variable times under the unknown dynamics, and local observables are measured (e.g., via randomized measurements) to yield time-resolved traces of few-body correlation functions, containing information about the implemented dynamics. This information is compiled, in the first step of the protocol, to determine the dynamics using Hamiltonian and Lindbladian Learning. This yields a set of "optimal" Hamiltonian and Lindbladian parameters (optimality specified below), along with their associated uncertainties. Using the quantum simulator as a computational device to compute observable expectation values, $\expval{O_t}$, we need to bound the uncertainty in $\expval{O_t}$ resulting from our uncertainty in the Hamiltonian and Lindbladian. Equivalently, we must ensure that the error incurred by assuming the device implements the learned Hamiltonian and Lindbladian does not exceed a specified threshold, with high confidence constituting step 2 of our protocol.}
    \label{fig:overview}
\end{figure*}

We present an overview of the core principles underlying quantum simulation with bounded error, based on Hamiltonian and Lindbladian Learning. Detailed derivations and proofs are provided in Appendix~\ref{app:theory}. Our primary focus is on analog quantum simulation in the NISQ regime, with particular emphasis on trapped-ion platforms, for which we report experimental realizations. While the discussion begins with the idealized setting of isolated quantum systems governed by Hamiltonian dynamics, the framework is subsequently generalized to encompass weakly dissipative open-system dynamics.

\subsection{Analog Quantum Simulation}\label{subsec:analogQS}

In analog quantum simulation a well controllable quantum many-body system is utilized to mimic a system of interest. Prime candidates of quantum simulators are systems of trapped ions, Rydberg atom tweezer arrays, superconducting qubits representing spin models, and ultracold bosonic and fermionic atoms in optical lattices representing Hubbard models~\cite{Blatt2012,Gross2017,Altman2021,Monroe2021,Fauseweh2024,Browaeys2020,Adler2024,Haghshenas2025,Manovitz2025,Phasecraft2025,Phasecraft22025,Mark2025,Abanin2025}.

Here, we focus on non-equilibrium (quench) dynamics which implements the unitary time evolution $U_{t}=e^{-iHt}$ for some
initial state $\rho_{0}$ and a time independent Hamiltonian, $H$, yielding the time-evolved state $\rho_{t}=e^{-iHt}\rho_{0}e^{iHt}$. 
We are interested in the situation  where the \emph{computational task} to be performed by the quantum simulator is to evaluate the expectation values of some observable $O$. We use the following notation to denote this computational task (forward model),
\begin{equation}
    H\;\mapsto\;\langle O_{t}\rangle=\mathrm{Tr}[O\rho_{t}]=\mathrm{Tr}[Oe^{-iHt}\rho_{0}e^{iHt}]. \label{eq:Hmap}
\end{equation}

The Hamiltonians which analog quantum simulators are supposed to realize should be understood as {\em effective} Hamiltonians that emerge within the low-energy subspace of an engineered composite many-body system subject to external control fields. While platforms like trapped ions and neutral atoms have well-established theoretical derivations for these effective Hamiltonians (see also below), the actual Hamiltonian realized experimentally is not known a priori and must be inferred from data subject to statistical uncertainty. Crucially, this uncertainty affects both the numerical values of Hamiltonian parameters and its operator content. The latter referring to the elements of a certain operator basis, such as the Pauli operators in spin models, which occur in the expansion of the Hamiltonian in this basis. Note that the experimentally realized Hamiltonian might contain for instance higher-order interactions which are absent in the theoretical model Hamiltonian. Thus, Hamiltonian learning becomes a multi-parameter estimation problem: inferring unknown coupling constants and determining, within a finite measurement budget, the operator content of the Hamiltonian.

\subsection{Trapped-Ion Quantum Simulator}\label{sec:TrappedIonQS}

Trapped ions forming one- or two-dimensional crystals in electromagnetic traps can be turned into an interacting quantum many-body system by dressing two of their long-lived electronic levels with laser light coupling their electronic and motional states~\cite{Blatt2012,Monroe2021}. This trapped ion platform provides the basis for experimental work discussed below (see Sec.~\ref{sec:ExpResults} and App.~\ref{app:exp}). 

An {\em effective spin-$1/2$ Hamiltonian}  can be  derived using adiabatic elimination of the phonon degrees of freedom as second-order perturbative expansion in the Lamb–Dicke parameter. The resulting Hamiltonian takes the form of a long-range Ising model \cite{Porras2004,Monroe2021}, 
\begin{equation}\label{eq:Ising}
H_{I}=\sum_{i<j}J_{ij}\,\sigma_{i}^{x}\sigma_{j}^{x},
\end{equation}
where an additional transverse-field $H_B=B\sum_i\sigma^z_i$ can be engineered in a suitable interaction picture by shifting the center frequency of the bichromatic laser field coupling the qubits. Here, and in the following, $\sigma_{i}^{x,y,z}$ denote the Pauli matrices, and the spin–spin
couplings $J_{ij}$ are mediated by the collective motional modes
of the trap. The couplings decay approximately as $J_{ij}\sim J_{0}/|i-j|^{\alpha}$
with the distance between ion $i$ and $j$, and $0<\alpha<3$.
Note that higher-order (many-body) corrections
to the above effective description will always be present, but are expected
to remain small in a manner that will be defined operationally (see
below).

In the experiments, we will be mainly interested in the regime of strong transverse field \( B\gg J_{0} \), where the above Hamiltonian reduces to a long-range \( XY \) model \cite{Jurcevic2014},
\begin{equation}\label{eq:XYmodel}
H_{XY} = \frac{1}{2}\sum_{i<j} J_{ij}\,(\sigma_i^x \sigma_j^x + \sigma_i^y \sigma_j^y).
\end{equation}

\subsection{Bounded-Error Quantum Simulation via Hamiltonian Learning}\label{sec:BEQStheory}

The unavoidable uncertainty in determining the  Hamiltonian $H$ with a finite measurement budget propagates to an uncertainty in the output of the computation $H \mapsto \langle O_t \rangle$. Hence, we need to develop protocols addressing this inverse problem that are both experimentally feasible and theoretically rigorous, capable of (i) 
experimentally learning $H$ with quantified
uncertainties, and (ii) establish rigorous error bounds
on the predicted expectation values $\langle O_{t}\rangle$, given
the experimentally accessible characterization of $H$. 
Crucially, such protocols must remain valid in regimes of quantum advantage, where the computation $H\;\mapsto\;\langle O_{t}\rangle$
may exceed classical computational capabilities. 

We now present a two-step protocol that realizes this concept within an experimentally accessible setting, see Fig.~\ref{fig:overview} for an overview of the protocol. 

\begin{figure*}
    \centering
    \includegraphics[width=1\linewidth]{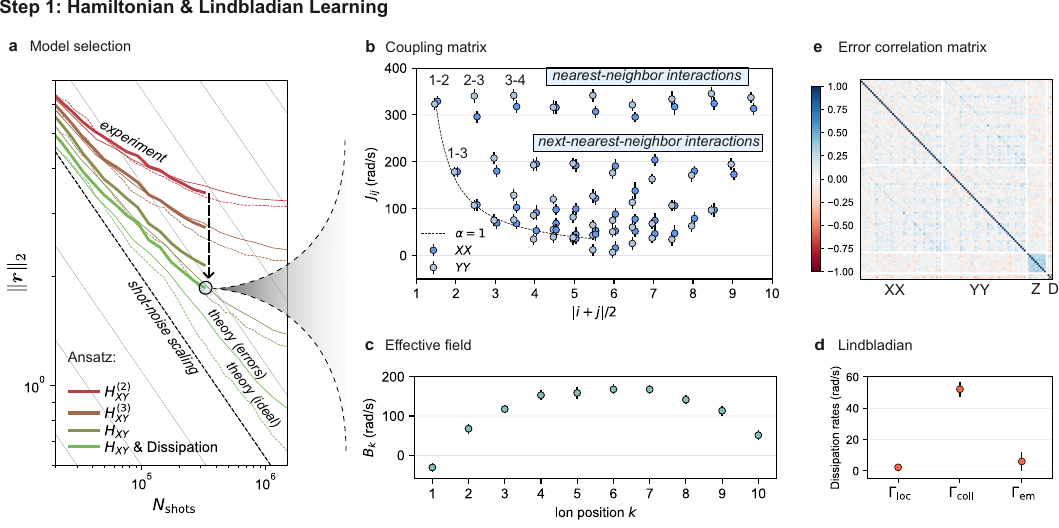}
    \caption{ {\bf Step 1 of the BEQS protocol: Hamiltonian and Lindbladian Learning.}
    ({\bf a}) 
    In an experiment with $N=10$ ions, we monitor the residual norm, $\Vert \vec{r}\Vert_2$, representing the \emph{learning error}, as a function of $N_{\rm shots}$ for different ansätze $H_{XY}^{(k)}$ (Eq.~\eqref{eq:ansatzK}) with increasing maximum interaction distance $k$, with and without dissipation (thick lines). Ansätze with restricted interaction range show strong deviations from the ideal shot-noise scaling, allowing us to detect inadequately chosen operator ansätze for the Hamiltonian and Lindbladian~\cite{Olsacher2025}. The ansatz closest to the anticipated shot-noise scaling consists of a long-range XY model, Eq.~\eqref{eq:XYmodel}, with additional transverse field, $H=H_{XY}+\sum_i B_i\sigma^z_i$, and a Lindbladian dominated by collective dephasing, $L_{\rm coll}=\sum_i\sigma_i^z$. The remaining deviation from shot-noise scaling is attributed to measurement errors (see App.~\ref{app:exp}). To verify that, we re-simulate the experiment with and without measurement errors (thin solid/dashed lines), and reproduce the observed and ideal (shot-noise) scaling, as well as a deviation from shot-noise scaling if measurement errors are included.
    The learned Hamiltonian parameters including $1\sigma$ error-bars are shown in ({\bf b}) and ({\bf c}). 
    In ({\bf b}) we show the interaction matrices, $J_{ij}^{(x,x)}$ and $J_{ij}^{(y,y)}$. Each interaction value is plotted at the midpoint $|i+j|/2$ between involved sites. This collapses the $(i,j)$ plane while preserving physically relevant structure: translation invariance appears as horizontal arrangements and spatial decay of interactions manifests as a systematic decrease in magnitude as $|i-j|$ increases. In this representation, points at the same horizontal location correspond to interactions centered around the same center point, $|i+j|/2$ , while their vertical spread reflects how the interaction strength varies with separation. This enables quick visual identification of whether the system is homogeneous, long-ranged, or exhibits spatial modulation of couplings.
    In ({\bf c}) we show an effective site-dependent magnetic fields $B_i$, and $({\bf d})$ shows the learned dissipation rates for local and collective dephasing, as well as spontaneous emission.
    Error bars are obtained via $N_b=300$ bootstrap resamples of the measurement outcomes, which also yields the error covariance matrix $\Sigma$ (Eq.~\ref{eq:HDistribution}), represented here as a \emph{correlation matrix} $\hat\Sigma=\mathrm{diag}(\Sigma)^{-1/2}\,\Sigma\,\mathrm{diag}(\Sigma)^{-1/2}$ in ({\bf e}), showing only weak off-diagonal correlations. For a detailed discussion of these results see Sec.~\ref{subsec:ExpResultsIntegral}.
    }
    \label{fig:Hlearning10}
\end{figure*}

 \subsubsection*{Step 1: Characterization via Hamiltonian Learning}
 \label{sec:step1H}

Physical many-body Hamiltonians are expected to
contain only a polynomial number of local terms acting on few sites. 
Hamiltonian learning exploits this structure to infer the operator content and the parameters  of $H$ from experimentally accessible few-body correlation data, 
providing a sample-efficient alternative to full process tomography. 
Such data can be obtained using established techniques including randomized measurements~\cite{Elben2022} 
or overlapping tomography~\cite{Cotler2020,Hansenne2025,Wei2024}. 

The starting point is to parametrize a class of local Hamiltonians via an ansatz
\begin{equation}
    \label{eq:ansatzH}
    H(\vec{c}) = \sum_{m=1}^{M} c_m h_m,
\end{equation}
where $\{h_m\}$ form a physically motivated operator basis 
[e.g., Pauli strings for spin models as in Eqs.~\eqref{eq:Ising} and~\eqref{eq:XYmodel}]. 

Hamiltonian learning protocols proceed by preparing a family of initial states on the quantum device, and monitoring a set of correlation functions for various quench times $t$. These experimental data provide a set of constraints for the Hamiltonian ansatz in Eq.~\eqref{eq:ansatzH}, taking the form of linear equations in $\vec{c}$. Within the class of ansatz Hamiltonians, the optimal coefficients $\vec{c}^*$ are obtained by minimization of an appropriate cost function,  
\begin{equation}
\label{eq:cost_function}
\vec{c}^{\,*}
= \arg\min_{\vec{c}} \Vert \tilde{\vec{M}}\vec{c} - \tilde{\vec{b}} \Vert_2^2,
\end{equation}
with matrix $ \tilde{\vec{M}}$ and vector $\tilde{\vec{b}}$ determined from experimental data. For instance, they can be obtained from estimating derivatives~\cite{Franca2022} or integrals~\cite{Olsacher2025} of time-traces of few-body correlation functions (see Appendix~\ref{app:theory} for explicit formulas).

A central diagnostic for model selection, i.e., choosing an appropriate and sufficiently expressive operator ansatz in Eq.~\eqref{eq:ansatzH}, \emph{and} the detection of systematic errors is the scaling of the residual norm, $\Vert\vec{r}\Vert_2 = \Vert \tilde{\vec{M}}\vec{c}^* - \tilde{\vec{b}} \Vert_2$, with the number of experimental runs, $N_{\rm shots}$. This quantity represents the \emph{learning error}---under a valid and sufficiently expressive Hamiltonian ansatz and negligible systematic errors, such as measurement errors and readout noise, one expects the residual norm to exhibit an asymptotic $1/\sqrt{N_{\rm shots}}$ decay. In contrast, systematic deviations from the assumed model manifest as a saturation of the residual norm at a finite $N_{\rm shots}$, i.e., the emergence of a plateau~\cite{Olsacher2025}. The onset of such a plateau indicates that the operator structure of the ansatz Hamiltonian is insufficient to capture the data. It may also reveal violations of underlying model assumptions, for example when dissipative dynamics have not been accounted for (see Sec.~\ref{sec:Open} for an extension to Lindbladians), when the assumption of a Markovian master equation is invalid—e.g., due to non-Markovian effects—or due to the presence of measurement errors and readout noise. In Fig.~\ref{fig:Hlearning10}a, we show the emergence of such plateaus in an experiment with $N=10$ ions, allowing us to detect missing ansatz terms and small measurement/readout errors. A more detailed discussion on the possibility to detect systematic errors is provided in Appendix~\ref{app:systematic_errors}.

When data are obtained via randomized measurements, this model-selection procedure can be carried out entirely via classical postprocessing. The ansatz is iteratively refined for fixed $N_{\rm shots}$ through reparameterization and regularization steps, as detailed in Appendix~\ref{app:HLLL}. These steps can significantly decrease the number of parameters $M$, which is especially relevant for large system sizes or when higher-order interaction terms are included. Rather than learning every individual coefficient, effective symmetries, such as translation invariance, or spatial averaging over many weak $N$-body terms can be incorporated into the regularization to constrain the ansatz to a small number of physically motivated parameters, offering a route to extend the protocol to regimes where resolving every higher-order term individually would otherwise become impractical.

In view of the finite measurement budget, the learning procedure yields not a single estimate of Hamiltonian parameters, but a distribution over parameters consistent with the data. Within the regime of sufficiently many measurement shots, we expect this distribution to be well-approximated by a Gaussian distribution (see Appendix~\ref{sec:error_models} for a justification), giving rise to the following ensemble of Hamiltonians,
\begin{eqnarray}
\label{eq:HDistribution}
H'(\vec{g}) &=& H_{\mathrm{learned}} + V(\vec{g}) \\
&\equiv& \sum_{m=1}^{M} \left(c_m^* + g_m\right) h_m,
\qquad 
\vec{g} \sim \mathcal{N}(0,\Sigma). \nonumber
\end{eqnarray}
The covariance matrix $\Sigma_{mn}=\mathbb{E}[g_m g_n]$, with $\mathbb{E}$ denoting the average over $\vec{g}\sim\mathcal{N}(0,\Sigma)$, captures statistical uncertainties (and their correlations) in the learned ansatz parameters $c_m^\ast$ arising from finite measurements. In practice, it can be estimated using nonparametric resampling techniques such as bootstrap or jackknife (described below), but also the standard Bayesian approach can be considered~\cite{Evans2019}. While ensembles of the form of Eq.~(\ref{eq:HDistribution}) have appeared previously as phenomenological noise models~\cite{Cai2024}, here they encode the inferred posterior uncertainty of the effective Hamiltonian.

Let us now discuss possible extensions of Eq.~(\ref{eq:HDistribution}). The ensemble can be extended by introducing additional ansatz terms and corresponding uncertainties, and extensions beyond the Gaussian approximation can be employed when the distribution is non-Gaussian~\cite{Cai2024}. Equation~(\ref{eq:HDistribution}), however, captures the statistical uncertainty of the learned parameters only within the chosen model class. 
It does not account for possible sources of systematic error in an experiment. 
In particular, additional bias may arise from limitations of the chosen operator ansatz, from regularization or structural constraints imposed in the learning procedure, and from imperfections in measurement and readout. In addition, systematic bias may arise from misspecification of the dynamical model itself, for example if the assumed Lindbladian description does not fully capture the underlying system dynamics. In practice, additional sources of systematic error can be diagnosed within our framework through deviations from the expected shot-noise scaling of the residual norm, as discussed above. A more detailed discussion including examples of systematic errors can be found in Appendix~\ref{app:systematic_errors}.

Such contributions are not contained in the covariance matrix $\Sigma$ underlying Eq.~(\ref{eq:HDistribution}), but may in principle be incorporated through extended models. For instance, systematic errors leading to a systematic bias in the learned parameters can be described by a nonzero mean offset, $g\sim\mathcal{N}(\vec{k},\Sigma)$, where $\vec{k}$ quantifies the resulting bias. However, observing that $\|\vec{r}\|_2\propto 1/\sqrt{N_{\mathrm{shots}}}$ indicates that $k_i\ll \sqrt{\Sigma_{ii}}$, meaning that parameter uncertainties are much larger than any potential systematic biases~\cite{Olsacher2025}, thereby justifying the ensemble in Eq.~(\ref{eq:HDistribution}).

Experimental results implementing this learning protocol for ten ions are shown in Fig. ~\ref{fig:Hlearning10}, and will be discussed in detail in Sec.~\ref{sec:ExpResults}. 

\begin{figure*}
    \centering
    \includegraphics[width=1\linewidth]{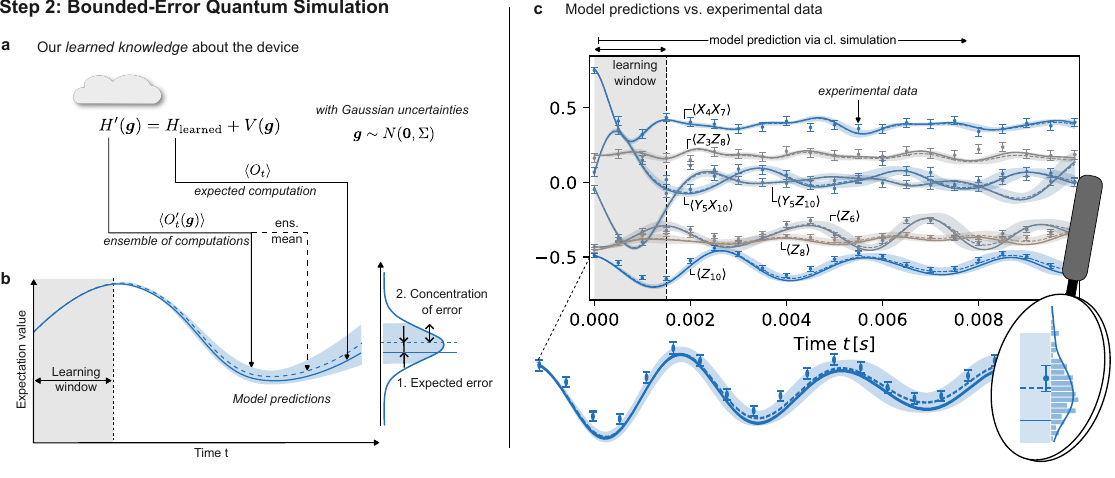}
    \caption{{\bf Step 2 of the BEQS protocol: Error bounds.} ({\bf a}) Following Hamiltonian and Lindbladian Learning, our knowledge of the realized dynamics is described by an ensemble of Hamiltonians~\eqref{eq:HDistribution} (and Lindbladians~\eqref{eq:statistical_model}), parametrized by a random Gaussian variable, $\vec{g}$. While the mean learned Hamiltonian $H_{\rm learned}$ would realize the expected computation of an observable expectation value, $\expval{O_t}$, the uncertainty in $H'(\vec{g})$ ultimately propagates non-linearly into observable expectation values. ({\bf b}) In a BEQS we need to ensure that these uncertainties do not lead to large errors relative to the expected computation, $\expval{O_t}$. Here, two errors are of interest (see Sec.~\ref{sec:ErrorBoundsShort}): the expected error, as the difference between $\expval{O_t}$, and the ensemble mean $\mathds{E}[\expval{O'_t(\vec{g})}]$, and the concentration of $\expval{O'_t(\vec{g})}$ around the ensemble mean, which can be controlled by a concentration inequality, effectively bounding the probability of large deviations. ({\bf c}) For the learned Hamiltonian and Lindbladian ensemble in Fig.~\ref{fig:Hlearning10}, we can still classically simulate the dynamics of a state under the ensemble. Solid lines represent trajectories predicted by the mean Hamiltonian and Lindbladian. Shaded areas represent the $95\%$ prediction intervals estimated from empirical percentiles using $50$ samples from the Hamiltonian and Lindbladian ensemble. Dashed lines represent the respective estimates of the distribution means. From these simulations one could estimate the expected error, Eq.~\eqref{eq:ExpectedError}, and a concentration bound, Eq.~\eqref{eq:TailBound}, including corresponding uncertainties through bootstrapping. For more details see Secs.~\ref{sec:ErrorBoundsShort} and Appendix~\ref{sec:boundedError}.}
    \label{fig:BoundedError}
\end{figure*}

\subsubsection*{Step 2: Establishing Error Bounds}\label{sec:ErrorBoundsShort}

Given our knowledge about the experimentally realized dynamics, as modeled by the ensemble in Eq.~\eqref{eq:HDistribution}, the expectation value of an observable $O$ at time $t$ is not uniquely determined but is described by an ensemble
\begin{equation}
\label{eq:ensembleO}
\langle O_t'(\vec{g}) \rangle
= \mathrm{Tr}\!\left[ O\, e^{-i H'(\vec{g}) t}\, \rho_0\, e^{i H'(\vec{g}) t} \right],
\end{equation}
where $\rho_0$ denotes the initial state.

In the following, our objective is to quantify, and subsequently bound, the \emph{error} we are making, when \emph{expecting} the device to perform the computation $H_{\rm learned}\mapsto \expval{O_t}$, despite our uncertainty. To this end, we need to quantify the deviations between $\expval{O_t}$, and the ensemble-induced predictions in Eq.~\eqref{eq:ensembleO}. These deviations consist of two complementary components~\cite{Cai2024}:

(i) Requiring that deviations of $\langle O_t'(\vec{g}) \rangle$ from their ensemble mean remain below a threshold $\eta$ with probability $p$ can be expressed through a concentration bound of the form
\begin{equation}
\Pr\!\left(\bigl|\langle O_t'(\vec{g}) \rangle- \mathds{E}\!\left[\langle O_t'(\vec{g}) \rangle\right]\bigr|\le \eta\right)\ge p(\eta),
\end{equation}
or, stated differently
\begin{equation}
\label{eq:TailBound}
\bigl|\langle O_t'(\vec{g}) \rangle- \mathds{E}\!\left[\langle O_t'(\vec{g}) \rangle\right]\bigr|\le \eta(p),
\end{equation}
with probability at least $p$.

(ii) Due to the non-linear dependence of Eq.~\eqref{eq:ensembleO} on $\vec{g}$, the ensemble mean, $\mathds{E}\!\left[\langle O_t'(\vec{g})\rangle\right]$, does not, in general, coincide with the value predicted by the mean Hamiltonian, $\expval{O_t}=\expval{O'_t(\vec{0})}$. 
The resulting bias between $\langle O_t\rangle$ and the ensemble mean, induced by uncertainty in the learned generator parameters, is quantified by the expected error
\begin{equation}
\label{eq:ExpectedError}
\bigl|
\langle O_t \rangle - \mathds{E}\!\left[\langle O_t'(\vec{g}) \rangle\right]
\bigr|.
\end{equation}
Together, Eqs.~\eqref{eq:TailBound} and~\eqref{eq:ExpectedError} fully describe the error as deviations between $\expval{O_t}$ and the ensemble $\langle O_t'(\vec{g}) \rangle$. In Figs.~\ref{fig:BoundedError}a and b, we show a schematic illustration of the above errors. The expected error quantifies the bias induced by parameter uncertainties when propagating predictions to observables. We emphasize that this quantity does \emph{not} represent the total experimental systematic error, but rather the component arising from uncertainty in the learned generator. In case of additional systematic errors, extended ensembles can be considered (see discussion below Eq.~\eqref{eq:HDistribution}). At the end of this section, we will present additional error bounds resulting from such extended ensembles.

When classical simulation is feasible, both errors can be determined by numerically propagating the learned Hamiltonian ensemble, giving access to the expected error and $\eta(p)$ as a prediction (or uncertainty) interval (see Fig.~\ref{fig:BoundedError}c for an example using the experimentally learned ten-qubit Hamiltonian and Lindbladian). In regimes of quantum-advantage, we need to compute analytic bounds on the expected error and expressions for $\eta(p)$. We address this problem through two complementary approaches.

In the first approach we use bounds adapted from Ref.~\cite{Cai2024}, which shows that under an uncorrelated Gaussian error model with uniform variance $\delta^2$, Eq.~\eqref{eq:TailBound} is bounded by
\begin{equation}\label{eq:bound_deviation}
    2t\delta\sqrt{M}c\norm{O},
\end{equation}
with probability at least $p=1-2e^{-c^2/2}$, and the expected error in Eq.~\eqref{eq:ExpectedError} is bounded by
\begin{equation}\label{eq:bound_expected}
    [e^{{2t^2\delta^2 M}}-1]\norm{O},
\end{equation}
where $\norm{O}$ denotes the operator norm. Both bounds can be directly obtained from the learned Hamiltonian ensemble and provide non-trivial bounds for time-scales where $\sqrt{M}t\delta\leq \mathcal{O}(1)$. As typically $\delta\propto 1/\sqrt{N_{\rm shots}}$ the measurement budget to reach a given target accuracy at time $t$ is at most polynomial in time and number of Hamiltonian parameters $M$. We note that in the context of Hamiltonian and Lindbladian Learning, more complicated error models can occur with non-uniform variance or correlated errors with non-diagonal covariance $\Sigma$. The resulting error bounds are discussed in Appendix~\ref{sec:long}.

For local Hamiltonians, the above bounds depend only on uncertainties within an observable’s past light cone, defined by sites within $d\leq v_{LR}t$, where $v_{LR}$ is the Lieb-Robinson (LR) velocity~\cite{Anthony2023}. Since the light cone grows linearly in time, the bounds remain independent of system size. Hamiltonians with long-range interaction tails, such as the experimental Hamiltonian in Eq.~\ref{eq:XYmodel}, require special treatment, as standard LR bounds do not apply. Taking these aspects into account, we present generalizations of the resulting error bounds in Appendix~\ref{sec:long}.

In the second approach we consider short evolution times, where $e^{-iH'(\vec{g})t}\approx \id - iH't -\frac{1}{2}(H't)^2$. We use this second-order expansion to obtain analytic bounds of the form
derive a short-time expansion of the fluctuation and apply a sub-Gaussian bound to the linear term and a 
Hanson–Wright inequality to the quadratic term. Combining the two via a union bound yields the following high-probability 
envelope
\begin{equation}\label{eq:HW_bound}
\bigl|\mathds{E}\left[\expval{O'_{t}(\vec{g})}\right]-\expval{O'_{t}(\vec{g})}\bigr| 
\leq \alpha_{\rm GC}(\eta,t) t \delta + \alpha_{\rm HW}(\eta)\frac{t^2}{2}\delta^2,
\end{equation}
holding with probability at least $1-\eta$, where we have assumed an uncorrelated Gaussian error model with uniform variance $\delta^2$. The functional dependencies of the parameter $\alpha_{\rm GC}$ follow from standard Gaussian concentration formulas~\cite{Vershynin2018}, together with  the above second-order expansion. Similarly, the parameter $\alpha_{\rm HW}$ is derived from the Hanson-Wright inequality~\cite{Rudelson2013}, stating similar concentration bounds for quadratic forms in Gaussian random variables. Both parameters depend on the initial state $\rho_0$, and the observable $O$. Computing precise numerical values for both parameters requires only the expectation values of a few-body observables in the input state $\rho_0$, which is typically a product state. These parameters can therefore be evaluated efficiently, even in regimes of quantum advantage. Moreover, the bounds can be directly extended to sub-Gaussian noise models, whose tails decay at least as fast as the tails of a Gaussian. In this case $\delta^2$ is a variance proxy controlling how fast the tails decay. For explicit expressions for the above bound we refer to Appendix~\ref{sec:short}.

Finally, we emphasize that if systematic biases—e.g., due to measurement errors or readout noise—are present and can be characterized, one may consider more general ensembles with $\vec{g} \sim \mathcal{N}(\vec{k}, \Sigma)$, with $\vec{k}$ characterizing the bias (see discussion below Eq.~\eqref{eq:HDistribution}). Since the total error is, in the worst case, additive, the additional error due to the bias $\vec{k}$ can be bounded using Duhamel's formula~\cite{Cai2024}, yielding an additional error term
\begin{equation}\label{eq:BoundSystematic}
2t\norm{\vec{k}}_1\norm{O},
\end{equation}
where $\norm{\vec{k}}_1 = \sum_{i=1}^M |k_i|$ denotes the 1-norm (see Appendix~\ref{app:systematicErrorBound} for a proof). Again, for local Hamiltonians, these bounds can be improved to only depend on errors within an observable's past light cone, and thus become system size independent.

\subsection{Open-System and Lindbladian Learning}\label{sec:Open}

Experimental quantum simulators typically operate in a weakly dissipative regime~\cite{Olsacher2025}, where residual couplings to the environment and control noise induce decoherence during evolution. In this regime, instead of the unitary dynamics generated by $H$ the system dynamics are well described by a Lindblad master equation~\cite{Gardiner2004}
\begin{equation}\label{eq:dynamical_model}
     \frac{\mathrm{d}\rho}{\mathrm{d}t} = \mathcal{L}\rho 
     \equiv -i[H,\rho] + \mathcal{D}(\rho),
\end{equation}
where $H$ generates coherent dynamics and the dissipator $\mathcal{D}(\rho) = \sum_m \Gamma_m \left(L_m \rho L_m^\dagger - \tfrac{1}{2}\{L_m^\dagger L_m, \rho\}\right)$, with jump operators $L_m$, captures irreversible processes~\footnote{We emphasize the validity of a master equation description for noise correlation times $\tau_c \rightarrow 0$.}.
In this case, the mapping in Eq.~\eqref{eq:Hmap} generalizes to
\begin{equation}
\label{eq:Lmap}
\mathcal{L} \;\mapsto\;
\langle O_t \rangle
= \mathrm{tr}\!\left[ O\, e^{\mathcal{L}t}\rho_0 \right],
\end{equation}
so the relevant learned object is the generator $\mathcal{L}$. Our bounded-error framework naturally extends by replacing Eq.~\eqref{eq:HDistribution} with an ensemble of learned Lindbladians,
\begin{equation}
\label{eq:statistical_model}
\mathcal{L}'(\vec{g}) 
= \mathcal{L}_{\mathrm{learned}}
+ \mathcal{V}(\vec{g}),  
\qquad 
\vec{g} \sim \mathcal{N}(0,\Sigma),
\end{equation}
where the uncertainties in both Hamiltonian and dissipative parameters are encoded in $\vec{g}$. 

\subsection{Developing the Theory of BEQS}

This section has outlined the concepts and methodological framework underlying BEQS, providing the basis for the results presented in the main text. For completeness and analytical rigor, Appendix~\ref{app:theory} develops the full theoretical formulation and methodological details of BEQS. This includes a review of the Hamiltonian and Lindbladian Learning protocols---referred to as the Integral \cite{Olsacher2025} and Derivative methods \cite{Franca2022}---together with an explicit derivation of quantitative error bounds relevant to quantum-simulation performance. These theoretical developments provide the mathematical foundations and precision guarantees that underpin the experimental and numerical analyses discussed in this work.

\section{Experimental Results} \label{sec:ExpResults}

Here, we report the experimental realization of the BEQS framework on a trapped-ion quantum simulator. Towards this end, linear strings of up to $51$ $^{40}$Ca ions are confined in the anisotropic harmonic potential of a linear radiofrequency trap. A long-range spin-spin Hamiltonian coupling qubits encoded in a pair of Zeeman levels of the $S_{1/2}\leftrightarrow D_{5/2}$ quadrupole transition is realized by illuminating all ions with a bichromatic light field that off-resonantly excites the ground-state cooled transverse motional modes of the string with qubit-state dependent strength. Local coherent qubit control enabled by a steerable laser beam interacting with single ions is used for input state preparation and measurement of arbitrary correlation functions. For the latter, qubit states are rotated prior to spatially resolved fluorescence detection, which provides high-fidelity state measurements in the energy eigenbasis.

In addition to the desired spin-spin interactions, local $\sigma^z_i$ terms arise by spatially varying interactions of the ions with the electromagnetic field, which will have to be added to the Hamiltonians of Eqs.~(\ref{eq:Ising}) and (\ref{eq:XYmodel}). For further experimental details, see Appendix~\ref{app:exp} and Refs.~\cite{Kranzl2022,Joshi2022}.

Two complementary sets of experiments were performed. The first realizes the full BEQS protocol for a system of \(N = 10\) spins, a regime in which classical simulations remain tractable. This experiment provides a comprehensive demonstration of BEQS, including the characterization of error scaling and the identification of conserved quantities and decoherence-free subspaces. It further reveals several nontrivial features of the system, such as spatial variations of the effective field \(B_i\) (see Eq.~\eqref{eq:Ising}), an exponential rather than algebraic decay of correlations, and the dominance of collective dephasing, while showing no evidence of higher-order terms within the experimental measurement budget.

The second experiment explores the scalability of the method by implementing the BEQS protocol for a chain of \(N = 51\) ions. Here, the focus lies on demonstrating that the learning and validating steps of BEQS remain experimentally feasible and informative well beyond the classically simulable regime. This establishes the applicability of BEQS as a method to test and verify quantum simulations on larger quantum devices.

\subsection{Experiments with $10$ Ions}\label{subsec:ExpResultsIntegral}

In our first experiment, we prepared two initial product states which are evolved under the experimental Hamiltonian and Lindbladian for discrete times $t_r\in\{0,0.5,1,1.5\}$ms. At each time step we estimated (almost) all up to three-qubit Pauli correlation functions via randomized measurements in the Pauli basis (see Appendix~\ref{app:IntegralMethodExp} for concrete state and measurement parameters and settings). This resulted in a measurement budget of $3.2\times10^5$ shots. This data set forms the basis of the subsequent analysis. We begin by detailing the specific steps of the learning algorithm, using the method described in Appendix~\ref{subsec:10IonIntegral} to obtain the learned Hamiltonian and Lindbladian ensembles, shown in Fig.~\ref{fig:Hlearning10}a and b. We then discuss the features of these ensembles shown in Fig.~\ref{fig:Hlearning10}c--f and how their extracted parameters compare to our theoretical expectations.

\subsubsection{Hamiltonian and Lindbladian Learning with Integral Method}

Our analysis begins in Fig.~\ref{fig:Hlearning10}a where we compare different ansätze for the operator structure of the dominant Hamiltonian part. Specifically, we consider the following class of Hamiltonians
\begin{equation}\label{eq:ansatzK}
    H^{(k)}=\sum_{i<j:\,\abs{i-j}< k} \qty[J^{(x,x)}_{ij}\sigma^x_i\sigma^x_j + J^{(y,y)}_{ij}\sigma^y_i\sigma^y_j] + \sum_{i}B_i\sigma^z_i.
\end{equation}
Here, all parameters are considered to be independent, and the parameter $k$ controls the maximum interaction distance, i.e., for $k=2$ this is a nearest-neighbor model, while for $k=N-1$ this is a long-range model with all-to-all interactions. In Fig.~\ref{fig:Hlearning10}a we compare the scaling of the residual norm for this class of Hamiltonians. While Hamiltonians with $k=2,3$ begin to show a plateau, the best scaling is obtained for a long-range Hamiltonian, illustrating the utility of the scaling of the residual norm to witness model inadequacies (see Sec.~\ref{sec:BEQStheory}). Moreover, including subdominant parts of the dynamics, here as a Lindbladian with local and collective dephasing, and spontaneous emission we can further improve the scaling. As a final step of the learning procedure, we use regularization to reduce the effect of shot-noise on the learned Hamiltonian parameters. We refer the reader to~\cite{Supp} Sec.~III and Ref.~\cite{Olsacher2025} for details.

The residual deviation from the expected shot-noise scaling can be attributed to (i) the finite resolution of the integration and (ii) measurement errors. To verify this, we re-simulated the experiment using the learned Hamiltonian and Lindbladian shown in Fig.~\ref{fig:Hlearning10}c--e as our model. One simulation was performed under ideal conditions with finer time-trace resolution, while a second simulation replicated the experimental settings, including a measurement error, where measurement outcomes ``0'' are flipped with a probability of $2\%$. Both simulations used the same total measurement budget. Under ideal conditions, the expected shot-noise scaling is recovered (Fig.~\ref{fig:Hlearning10}a, dashed lines), whereas the simulation including measurement errors reproduces the scaling of the residual norm observed experimentally (Fig.~\ref{fig:Hlearning10}a, solid lines).

\begin{figure}[t]
    \centering
    \includegraphics[width=\linewidth]{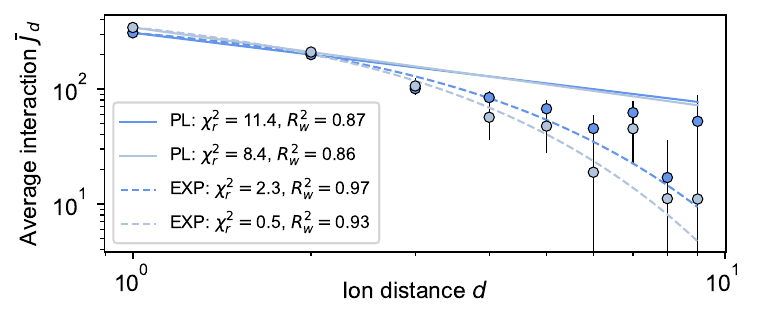}
    \caption{\textbf{Decay of interactions for $N=10$ ions.} Average interaction $\bar J_{i,i+d}$ as a function of the inter-ion distance, $d$. The decay is expected to approximately follow a power-law. One observes that the decay of interactions is faster than predicted by the ratio of nearest-neighbor to next-nearest-neighbor interactions. Here, an exponential decay provides better predictions for the long-range tails.}
    \label{fig:AlgDecay10}
\end{figure}

The learned parameters, including their uncertainties, are shown in Fig.~\ref{fig:Hlearning10}c-e. Several key features of the target model in Eq.~\eqref{eq:XYmodel} can be identified in the learned Hamiltonian:

{\it (i) Decay of the long-range interactions:} The long-range interactions of the Hamiltonian approximately follow a power-law decay $J_{ij}\approx J_0/\abs*{i-j}^\alpha$, see Sec.~\ref{sec:TrappedIonQS}. In Fig.~\ref{fig:AlgDecay10} we show the average interaction strength, $\bar J_d = 1/(N-d)\sum_i J_{i,i+d}$, as a function of the inter-ion distance $d$, for both, the interactions $\sigma_x\sigma_x$ and $\sigma_y\sigma_y$, which allows us to observe deviations from a power-law decay. We consider the decay of the nearest- and next-nearest-neighbor interactions and fit a power-law and an exponential decay law to these points. We observe that the decay estimated by a exponential decay law, $J_{ij}\propto e^{-\tau\abs{i-j}}$, with $\tau_x = 0.44 \pm 0.05$ and $\tau_y = 0.53 \pm 0.04$, provides a better predictions of the long-range tails compared to the power-law with $\alpha_x = 0.64 \pm 0.07$ and $\alpha_y = 0.7 \pm 0.1$. We note, that these deviations are well within the theoretical predictions of Eq.~\eqref{eq:JijTheory} and may vary with experimental parameters.

{\it (ii) Structure and size of interaction terms:} Fig.~\ref{fig:Hlearning10}b shows that, as expected, the interactions, $J_{i,j}^{(x,x)}$ and $J_{i,j}^{(y,y)}$, are very much homogeneous along the ion string. However, from theory calculations, we would expected nearest-neighbor interactions of $J_0\approx 288$ rad/s, which are slightly smaller than the experimentally measured values. This can be explained by the trapping potential's principal axes deviating by few degrees from the expected directions in the plane perpendicular to the ion string.

{\it (iii) Conservation of the total magnetization:} In Fig.~\ref{fig:Hlearning10}b we furthermore see that the Hamiltonian features a strong X--Y symmetry, i.e., $J_{i,j}^{(x,x)}\approx J_{i,j}^{(y,y)}$, for any pair of qubits $(i,j)$ giving rise to a conservation of the total magnetization, $J_z=\sum_k\sigma_k^z$, i.e., $[H_{\rm learned},J_z]\approx 0$.

{\it (iv) Local versus global dephasing:}
Figure~\ref{fig:Hlearning10}d shows that collective dephasing of the spins is the main source of decoherence. This type of dephasing originates from (fast) frequency fluctuations of the laser addressing all ions \footnote{We note that {\em fast} laser fluctuations appearing as collective dephasing can be modeled by a master equation. Instead, global magnetic field fluctuations $B$, which are typically {\em slow} on the experimental scales, will also give rise to a collective dephasing. However, this would be described as an inhomogeneous average of unitary dynamics with Hamiltonian $H(B)$ with $B$ static, averaged over a distribution $P(B)$.}. Other decoherence processes, such as local dephasing, which our method can distinguish from collective dephasing, and spontaneous emission remain subdominant. We expect a spontaneous emission rate of $\Gamma_{\rm em}\approx 2$ rad/s, which is well within error bars of the reconstructed Lindbladian. In the following we will demonstrate that our method cannot only distinguish local from collective dephasing, but can learn more general noise models from which collective dephasing emerges as the dominant contribution.

\paragraph*{Learning a complete dephasing model.} Fast stochastic fluctuations of laser phases and ambient magnetic fields during the time-evolution  lead to dephasing of the dynamics. This leads to a quantum master equation as in Eq.~\eqref{eq:dynamical_model} with dephasing Lindbladian~\cite{Gorini1976}
\begin{equation}\label{eq:general_dephasing}
    \mathcal{D}(\rho)=\sum_{i,j=1}^{N} \Gamma_{ij} (\sigma^z_i\rho \sigma^z_j-\frac{1}{2}\{\sigma^z_i\sigma^z_j,\rho\}).
\end{equation}
Note that the positivity condition, $\Gamma\succeq 0$, ensures that the generated evolution is completely-positive and trace-preserving (cptp). The above Lindbladian exhibits two limiting cases. Local (uncorrelated) dephasing leads to a diagonal matrix $\Gamma_{ij}=\Gamma_{\rm loc}\delta_{ij}$, whereas collective (correlated) dephasing results in a constant matrix $\Gamma_{ij} = \Gamma_{\rm coll}$ for all $i,j$. Diagonalizing the collective dephasing matrix transforms the master equation to Lindblad form with collective jump operator $L_{\rm coll}=\sum_i\sigma^z_i$. As mentioned before, 
in Figure~\ref{fig:Hlearning10}d we showed that the learning protocol can distinguish these two cases. In the following, we demonstrate that the full noise model in Eq.~\eqref{eq:general_dephasing}, i.e., the dephasing matrix $\Gamma$ itself, can be learned.

To this end, we first need to parametrize the upper triangular part of the dephasing matrix $\Gamma$, i.e., we define
\[
\Gamma(\vec{c}) =
\begin{bmatrix}
c_{M+1} & c_{M+2} & \cdots & c_{M+N} \\
& c_{M+N+1} & \cdots & c_{M+2N-1} \\
& & \ddots & \vdots \\
& & & c_{M'}
\end{bmatrix}.
\]
Then, the parameters $c_1,\dots,c_M$ parametrize the ansatz Hamiltonian, Eq.~\eqref{eq:ansatzH}, while the parameters $c_{M+1},\dots,c_{M'}$ parametrize the dephasing model. Note, that all remaining parameters of $\Gamma$ are determined by $\Gamma^{\top}=\Gamma$.

\begin{figure}[t]
    \centering
    \includegraphics[width=0.9\linewidth]{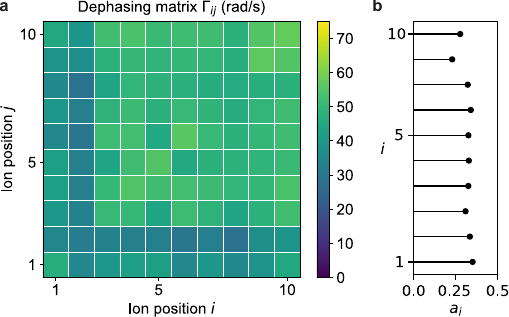}
    \caption{\textbf{Dephasing Lindbladian for $10$ ions.} ({\bf a}) Choosing a general dephasing Lindbladian as an ansatz, we can resolve the structure of the dephasing matrix $\Gamma$ in Eq.~\eqref{eq:general_dephasing}. While perfect collective dephasing leads to a constant dephasing matrix of the form $\Gamma_{ij}=\Gamma_{\rm col}$, independent single-qubit dephasing yields a diagonal dephasing matrix $\Gamma_{ij}=\Gamma_{\rm loc}\delta_{ij}$. In the experiment we can recover the structure of the dephasing matrix and transform the resulting master equation to Lindblad form by diagonalizing the dephasing matrix $\Gamma$. ({\bf b}) We find a dominant dephasing process, with jump operator $L=\sum_{i}a_i\sigma^z_i\approx L_{\rm coll}/\sqrt{N}$, resembling a collective dephasing of the qubits. Other jump operators are close to single- or few-qubit dephasing at much smaller rates.}
    \label{fig:dephasingmatrix}
\end{figure}

To ensure the reconstructed Lindbladian yields a CPTP evolution, instead of solving the least-squares problem in Eq.~\eqref{eq:cost_function}, we need to solve the following \emph{second-order cone program}~\cite{BoydVandenberghe2004}
\begin{align}\label{eq:gammapositive}
\vec{c}^{*} = \arg\min_{\vec{c}} \quad & \norm*{ \tilde{M} \vec{c} - \tilde{\vec{b}} }_2^2 \\
\text{s.t.} \quad & \Gamma(\vec{c}) \succeq 0.
\end{align}
Transforming the problem to \emph{epigraph form} and turning the resulting quadratic constraint $\|A c - b\|_2^2 \le t$ into a linear matrix inequality using the Schur complement one obtains a \emph{semidefinite program}, which can solved efficiently using \texttt{CVXPY}~\cite{Agrawal2018} and the solver \texttt{MOSEK}~\cite{mosek} (see App.~\ref{app:SDP}). The resulting dephasing matrix is shown in Fig.~\ref{fig:dephasingmatrix}a. 
Diagonalizing the dephasing matrix transforms the master equation to Lindblad form. In its spectral decomposition $\Gamma=\sum_m \Gamma_m\vec{a}_m\vec{a}_m^{\top}$, the eigenvectors define the corresponding jump operators $L_m=\sum_{i}a_i^{(m)} \sigma^z_i$ with associated rates $\Gamma_m$. For the learned dephasing matrix, the dominant eigenvector is shown in Fig.~\ref{fig:dephasingmatrix}b. It yields a jump operator that closely resembles the collective dephasing operator, $L_0=\sum_{i}a_i^{(0)}\sigma^z_i\approx L_{\rm coll}/\sqrt{N}$, with a corresponding rate of $\Gamma_0=460$ rad/s. The remaining jump operators correspond to single or few-qubit dephasing processes at significantly smaller rates of $23$ rad/s and below.

\subsubsection{Results for BEQS}
\begin{figure}[t!]
    \centering
    \includegraphics[width=.9\linewidth]{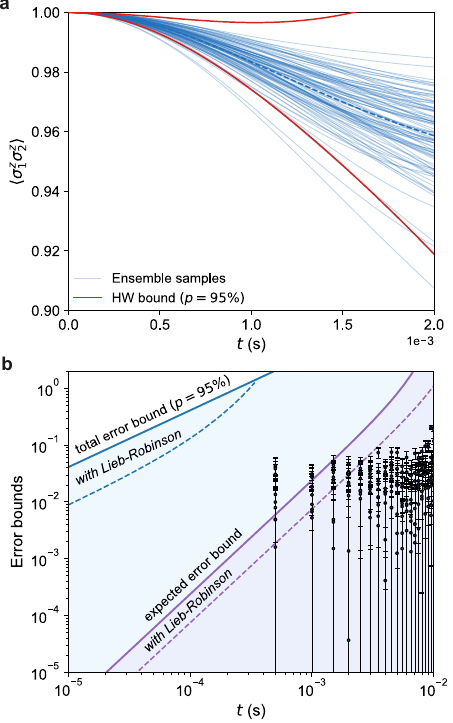}
    \caption{
    \textbf{Theory bounds for the ensemble of Fig.~\ref{fig:Hlearning10} with $N=10$ ions.} ({\bf a}) For short times we obtain bounds via Eq.~\eqref{eq:HW_bound} on a correlation function $\expval{\sigma^z_1\sigma^z_2}$, for an inital state $\ket{\psi}=\ket{0}^{\otimes N}$~\cite{figNote}, and Gaussian error (see Fig.~\ref{fig:Hlearning10}) with $\delta\approx 13$ rad/s. As a comparison, we show here $N_{\textrm{s}}=100$ simulated short-time trajectories, under Hamiltonians drawn randomly from the ensemble, which are well bounded by the Hanson-Wright envelope (red). (b) Analytical bounds on the expected error, Eq.~\eqref{eq:bound_expected} (violet), and the total error, i.e., the sum of bounds in Eq.~\eqref{eq:bound_expected} and Eq.~\eqref{eq:bound_deviation} (blue). For short times, a truncation of the Hamiltonian yields better bounds as the truncation error is much smaller than the error due to the uncertainty in the truncated Hamiltonian terms. Here, we consider a local observable at site $i=1$, and truncate the Hamiltonian at distance $R=3$. For comparison black markers indicate the true error, i.e., the difference between experimental data points and the predictions of the mean Hamiltonian and Lindbladian, including experimental shot-noise (error bars), obtained from exact numerical simulations.}
    \label{fig:error_bound_theory}
\end{figure}

We consider the learned Hamiltonian and Lindbladian model discussed in Fig.~\ref{fig:Hlearning10}~\footnote{We disregard terms that are zero within one standard deviation.}. The small system-size allows us to numerically simulate the complete dynamics of the learned ensemble which is shown for some few-body correlation functions in Fig.~\ref{fig:BoundedError}c. Here, the solid lines represent the expected computation of the device, $\expval{O_t}$, shaded regions represent $95\%$ prediction intervals, and dashed lines the ensemble mean of $\expval{O'_t(\vec{g})}$. These predictions can be compared with the actually measured experimental data represented by markers with $2\sigma$ error bars (compatible with the $95\%$ prediction intervals). While the vast majority of points overlaps with the simulation of the predicted model, some outliers remain, probably due to drifts of some control parameters or low-energy Langevin collisions with background gas molecules that may not be detected by the experiment control system.
Next, we will consider the short-time bounds based on the union bound inequality in Eq.~\eqref{eq:HW_bound}. 
To simplify the determination of the bound, we use the fact that we have identified a decoherence free subspace (see above), we will consider a state in that subspace as an example, namely $\ket{\Psi}=\ket{0}^{\otimes N}$. For this state we compute the bounds in Eq.~\eqref{eq:HW_bound} using the experimentally learned Hamiltonian, shown in Fig.~\ref{fig:Hlearning10}. The blue lines in Fig.~\ref{fig:error_bound_theory}a represent $N_{s}=100$ simulated trajectories under a randomly chosen Hamiltonian from the ensemble, and the dashed line represents the ensemble average. The bound obtained from Eq.~\eqref{eq:HW_bound} is shown in red plotted around the ensemble mean. Here, we have chosen $\eta=0.05$ ($95\%$ confidence).

Finally, we can obtain bounds directly from the learned error model in Fig.~\ref{fig:Hlearning10}e, i.e., the covariance matrix $\Sigma$. In Fig.~\ref{fig:error_bound_theory}b we show bounds on the expected error (violet lines), as well as a bound on the total error (blue lines), i.e., the sum of Eqs.~\eqref{eq:TailBound} and~\eqref{eq:ExpectedError}, relative to the expected computation. For short times, tighter bounds (dashed lines) are obtained under a truncation of the Hamiltonian, and estimating the truncation error via LR-bounds.

\subsection{Experiments with $51$ Ions}\label{subsec:51IonDifferential}
\begin{figure*}[t!]
    \centering
    \includegraphics[width=1\linewidth]{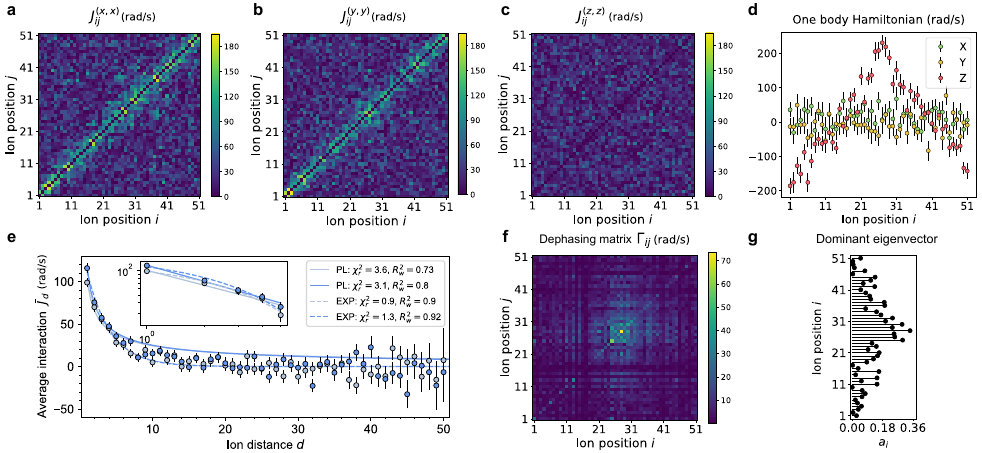}
    \caption{\textbf{Hamiltonian and Lindbladian learning with $N=51$ ions.} 
    We choose the Hamiltonian ansatz in Eq.~\eqref{eq:twobody_Lindbladian}, and the general dephasing model in Eq.~\eqref{eq:general_dephasing} as our ansatz for the Lindbladian, representing a total of $3N+9N(N-1)/2+N^2=14 \ 229$ parameters to be learned. The dominant interaction terms $J_{ij}^{(x,x)}$ and $J_{ij}^{(y,y)}$ are shown in ({\bf a}) and ({\bf b}) respectively. 
    Other interaction terms, such as $J_{ij}^{(z,z)}$ shown in ({\bf c}), are negligible with respect to our statistical error bars. In ({\bf d}) we show single-body interaction terms, where the $B_i^{(z)}$ fields are dominant, forming a triangular pattern form $-200$ to $+250$ rad/s, centered in the middle of the chain, while other fields remain small compared to their error bars.
    ({\bf e}) Averaged couplings $\bar J_d=1/(N-d)\sum_i J_{i,i+d}$ for $\sigma^x\sigma^x$ couplings (blue dots), and $\sigma^y\sigma^y$ couplings (gray dots),
    as a function of qubit distance $d=|i-j|$.
    The first five data points (shown in the inset plot in logarithmic scale) were fitted with a power-law decay model (PL) and an exponential decay model (EXP). We observe that the exponential decay model provides a better fit to the experimental data (with a higher $R_w^2$ score and smaller $\chi_r^2$ score) and therefore better explains the observed decay. The displayed error bars correspond to one standard deviation, and are estimated using the Jackknife (JK) leave-one-out method~\cite{Bergonzini1993}, where each JK sample is obtained by removing one of the $N_U$ settings.
    The dephasing matrix, $\Gamma_{ij}$ (Eq.~\eqref{eq:general_dephasing}), displayed in ({\bf f}), has most of its support concentrated near the center of the chain and features off-diagonal elements indicating collective dephasing mechanisms.
    Diagonalizing $\Gamma_{ij}$ transforms the master equation to Lindblad form. We identify a dominant dephasing process, with rate $\Gamma=364.5$ rad/s, associated with the jump operator
    $L=\sum_ia_i\sigma_i^z$. We show the spatial profile of the corresponding eigenvector in ({\bf g}). Other processes have smaller rates, below $125$ rad/s.
     }
    \label{fig:HL51} 
\end{figure*}
We now move on to the experimental protocol and results with $N=51$ ions.

\subsubsection{Lindbladian learning with the differential method}
Here, we use the method introduced
in Ref.~\cite{Franca2022}, and we assume that the Hamiltonian contains at most two-qubit Pauli operators
\begin{align}
\label{eq:twobody_Lindbladian}
H&=
\sum_i
\sum_{a=x,y,z}
B_i^{(a)}\sigma_i^{(a)}
+
\sum_{i<j}
\sum_{a,b=x,y,z}
J_{ij}^{(a,b)}\sigma_{i}^{(a)}\sigma_j^{(b)},
\end{align}
As shown in Refs.~\cite{Franca2022, Lam2026}, 
the sample complexity of the protocol scales as 
\mbox{$\mathcal{O}(3^{w_1+w_2}\log(N)\varepsilon^{-2})$},
where $w_1$ and $w_2$ denote the weights of the initial state and the measured operator, respectively, $N$ is the system size and $\varepsilon$ is the target precision. Therefore, the dependence on the system size is only logarithmic.
Increasing the locality $k$ of the Hamiltonian, to include higher-order interactions, such as three- or four-body terms, increases $w_1$ and $w_2$, according to the relation \mbox{$w_1+w_2=k+1$}. Consequently, the sample complexity required to achieve a given statistical precision also increases, while preserving the favorable logarithmic scaling with system size.

As for the experiment with $10$ ions, we model decoherence with the dephasing Lindbladian in Eq~\eqref{eq:general_dephasing}. 
It should be noted that originally the methods of Ref.~\cite{Franca2022} were developed for geometrically local or decaying interactions, whereas the Lindbladian in Eq~\eqref{eq:general_dephasing} has all-to-all interactions. However, in Sec.~II.B.2 of the Supplemental Material~\cite{Supp} we show that generators like that of Eq~\eqref{eq:general_dephasing} fit into the framework of Ref.~\cite{Franca2022}. In addition, in Sec.~II.B.1 we extend how to apply the methods of~\cite{Franca2022} to algebraically decaying interactions with a smaller exponent than those already discussed in the original work to ensure we cover the systems at hand. 

The method consists in using the Lindblad master equation in Eq.~\eqref{eq:dynamical_model} at $t=0$ to express the derivatives
\mbox{$b_k=\frac{\mathrm{d}}{\mathrm{d}t}\langle P_k \rangle_{\rho_k} \big|_{t=0}$}, with $P_k \in \{\sigma_x,\sigma_y,\sigma_z\}$ a Pauli observable, in terms of a linear system $\vec{b}=\vec{M}\vec{c}$, where $\vec{c}$ is the vector of Hamiltonian coefficients that are to be determined.
The entries $b_k$ are measured by interpolating the time traces $\expval{P_k(t)}$ with low degree polynomials and taking the degree one coefficient as the $b_k$ estimate.
The matrix $M$ is \textit{calculated} using simple algebra on $H$, $\mathcal{D}(\rho)$, $\rho_k$ and $P_k$; although we neglect the SPAM errors here, they can be incorporated---if characterized---following Ref.~\cite{Franca2022} (see Supplementary Note 9 therein), in which case the matrix $M$ will depend on the experimental data rather than being computed from Pauli algebra alone. Absent such a characterization, the protocol remains robust and retains its favorable sample complexity scaling provided SPAM errors are local and small relative to the target precision~\cite{Franca2022}. From our numerical tests, we found that, for the current experimental budget, the dominant source of error was the statistical uncertainty in estimating observables and their derivatives rather than the SPAM errors. We therefore did not include them explicitly, though we expect this to become necessary at larger measurement budgets, for which Ref.~\cite{Franca2022} provides a clear route to their incorporation.

We choose the initial states $\rho_k$ as \textit{partially mixed} states, i.e, in the form
\begin{equation}
    \rho_k = \rho_k^{(i)}\otimes\rho_k^{(j)}\otimes \left(\frac{\id}{2}\right)^{\otimes N-2},
    \label{eq:mixedstate}
\end{equation}
written up to the reordering of the qubits, and with $\rho_k^{(i)}\otimes\rho_k^{(j)}=|\psi_i\rangle\langle\psi_i|\otimes|\psi_j\rangle\langle\psi_j|$ two positive or negative eigenstates of Pauli operators.
Based on the decomposition of $H$ and $D(\rho)$, and the choice of initial states $\rho_k$ and chosen observables $P_k$, $M$ can be shown to \textit{sparse} and \emph{well-conditioned}, making it well-suited for solving the linear system using least-squares methods~\cite{Franca2022, Lam2026}. 
We measure $\mathbf{b}$ using randomized measurements: this involves $N_U$ random settings, where a random product state is prepared, time evolved with the system, and measured with $N_M$ shots in a basis generated by random single qubit rotations. The complete acquisition and postprocessing procedures are described in App.~\ref{subsec:10IonDiff}.
Moreover, as with the integral method, we can include regularization models, and impose additional constraints, such as positivity on the dephasing matrix $\Gamma$.

\subsubsection{Experimental Results}\label{subsec:H51}
The experiment was conducted with $N_U=200$ random settings, $N_M=200$ shots, and $N_t=11$ equally spaced time evolutions between $0$ and $1$ ms. This corresponds in total to $4.4 \times 10^5$ single-shot measurements.
We obtained a vector $\boldsymbol{b}$ of length $196 \ 908$ and a matrix $M$ of size $(196 \ 908,14 \ 229)$, that we constructed using the \texttt{scipy.sparse} format.
We considered polynomials of degrees $2$ to extract the derivatives $b_k$, and solved the system using \texttt{CVXPY}~\cite{Agrawal2018} and \texttt{MOSEK}~\cite{mosek} to impose the positivity constraint, $\Gamma\succeq 0$, as in Eq.~\eqref{eq:gammapositive}.

Our experimental results are shown in Fig.~\ref{fig:HL51}. In panels (a) and  (b), we show the two main contributions to the interaction Hamiltonian $J_{ij}^{(x,x)}$ and $J_{ij}^{(y,y)}$, while in panel (c), we show an example of the other two body terms, like $J_{ij}^{(z,z)}$, that can be neglected with respect to our statistical error bars. 
The single-body terms are shown in panel (d). The fields $B^{(z)}_i$ display an inhomogeneous profile, similar to the $10$ qubit results in Fig.~\ref{fig:Hlearning10}, with the values varying from $-200$ to $+250$ rad/s.

Having identified the main two body Hamiltonian contributions $J_{ij}^{(x,x)}$ and $J_{ij}^{(y,y)}$, we analyze their average coupling strengths $J_d$, to access the overall interaction trend as a function of the distance $d$, that we display in Fig.~\ref{fig:HL51}e.
While a power-law fit accurately captures the interacting decay
at small distances $d\le 5$, we observe that an exponentially decaying model, $J_d = J_0' e^{-\tau d}$, captures better the long range data $d> 5$. For $J_{ij}^{(x,x)}$ (blue dots), we obtained $J_0'=130.2 \pm 6$ rad/s and $\tau=0.28\pm 0.02$  and for $J_{ij}^{(y,y)}$ (gray dots), $J_0'=152.2 \pm 10.3$ rad/s and $\tau=0.31 \pm 0.02$.
Finally, we show in Fig.~\ref{fig:HL51}f the dephasing matrix $\Gamma_{ij}$, where the presence of non-zero off-diagonal coefficients reveals the presence of collective dephasing. In particular, in the center region of the ion chain, we see that the coefficient magnitudes are concentrated around the qubits $25$ to $32$. This is more visible looking at coefficients of the associated jump operator associated with the biggest eigenvalue of $\Gamma$, shown in panel (g). This indicates stronger fluctuations of the magnetic field towards the center of the chain.

\begin{figure}[t]
    \centering
    \includegraphics[width=\linewidth]{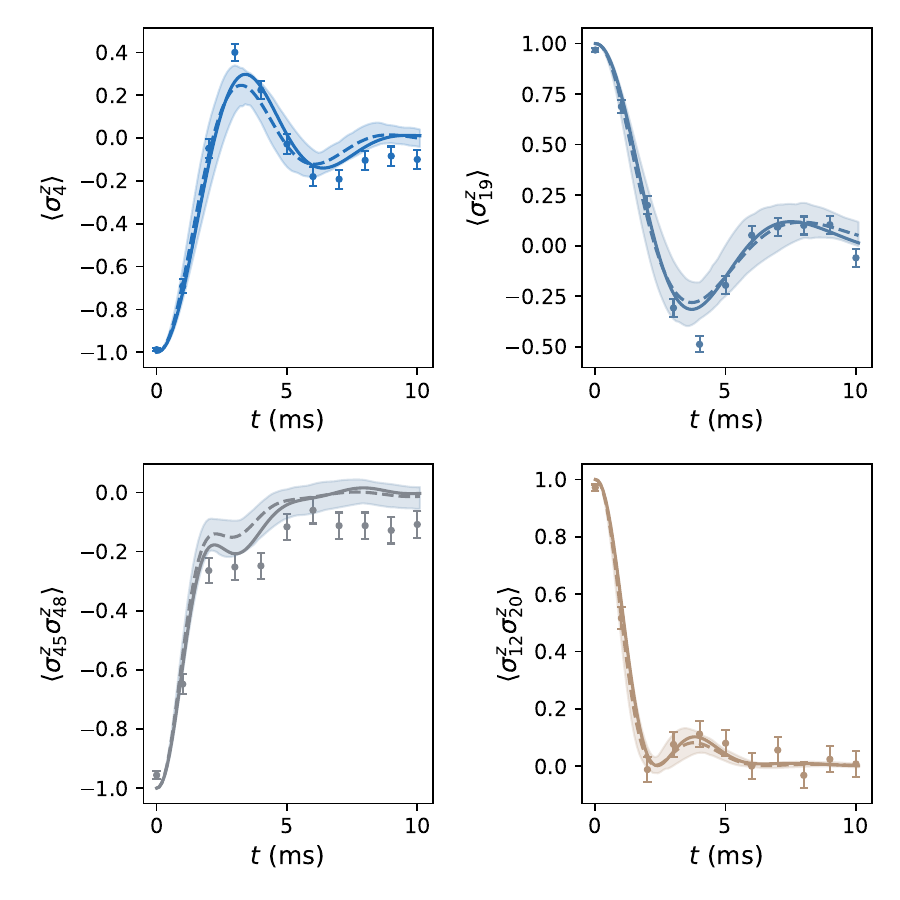}
    \caption{
\textbf{BEQS on the system with $N=51$ ions.}
For an initial N\'eel state, using the ensemble of Hamiltonians constructed from the learned mean and errorbars, we show expectation on measurements of some typical operators $\langle O_t'(\vec{g}) \rangle$ using the shaded regions. As explained in the Sec.~\ref{sec:BEQStheory} we observe differences between the dashed curves (the mean $\mathbb E\langle O_t'(\vec{g}) \rangle$) and the solid curves ($\langle O_t \rangle$ from the learned Hamiltonian).
The experimental data—dots and error bars—obtained in an independent run of the experiment on a different day from the learning-data acquisition corroborate the predicted uncertainty intervals, demonstrating consistency between the learned generator and the long-time quench dynamics. The TDVP simulation is done using Trotter time step $\delta t = 0.1$ ms and restricting maximum bond dimension to $\chi=100$. Numerically, the one and two point observables converge well using these parameters.
}
\label{fig:TheoryVsExp51}
\end{figure}

\subsubsection{BEQS with $N=51$ Ions}
We now analyze bounded error quantum simulation based on the learned Lindbladian. We use classical simulations of Hamiltonian evolutions sampled from the learned Hamiltonian ensemble. To sample a faithful dynamics, instead of directly using the measured Lindbladian parameters which have large error bars (Fig.~\ref{fig:HL51}), we use a regularized Lindbladian, see in the Appendix~\ref{app:HLLL}

In  the Fig.~\ref{fig:TheoryVsExp51} we show the predicted error bounds of the observable as a function of time. For the demonstration, we have taken N\'eel state $\ket{\psi}= \ket{\downarrow\uparrow\downarrow\dots}$ as the initial state and evolve it with the sampled Hamiltonians via time-dependent variation principle (TDVP)  in the tensor network representation using the TeNPy library \cite{10.21468/SciPostPhysLectNotes.5}. The expectation on $\langle O'_t(\vec{g}) \rangle$ (we choose some typical operators as seen in the figure) using a computational device is shown as the estimated $95\%$ prediction intervals on 67 Hamiltonian samples. Similar to the presentation for $N=10$ ions in Fig.~\ref{fig:BoundedError}c, here also, the dashed curves show the mean $\mathbb E [\langle O'_t(\vec g) \rangle]$, and  the solid curves show the $\langle O_t\rangle$ from the learned Hamiltonian. We observe the bias as defined in the Eq.~\eqref{eq:ExpectedError}. 

We point out that, to obtain such error bounds, we assumed  the time evolution to be solely governed by the Hamiltonian part of the Lindbladian.  In other words, we assumed the existence of a decoherence free subspace despite the fact that the learned dephasing matrix was not purely global. In principle, we could use the data about the learned Lindbladian and the tools developed here to quantify the effect of any deviations from the simplistic error model of collective dephasing. The prediction from the learned Hamiltonian ensembles are cross-validated with an independent run of the experiment, consisting in measuring the observables as functions of time, shown using dots and errorbars.
We observe that the classically simulated error bounds are mostly compatible with the observed data, with some discrepancies. We attribute these differences to  the  decoherence free subspace approximation as well as to inefficient Hamiltonian sampling due to large statistical errors in the learned Hamiltonian coefficients. 
Our study demonstrates that the framework of Lindbladian learning and BEQS can be faithfully applied in a large-scale scenario.

\section{Conclusions and Outlook}

We have developed a quantitative Hamiltonian–Lindbladian learning framework that reconstructs the generator of the dynamics together with certified uncertainties, and we have \emph{verified} the learned model by classically simulating the full error model and comparing its quantitative predictions with independent experimental measurements. This constitutes, to our knowledge, the first experimentally validated demonstration of Hamiltonian–Lindbladian learning, establishing that the reconstructed generator is both statistically well defined and predictive in the dynamical regime explored here. We have further shown that bounded-error quantum simulation can be performed using experimental data alone, providing in principle a path to regimes beyond classical simulability. Rigorous bounds of this type have been developed in the theoretical literature~\cite{Cai2024}, and our results supply the experimental interface required to evaluate them for a concrete analog quantum simulation experiment. As expected, these bounds are informative only over comparatively short evolution times. Because the present system size remains classically tractable, we have benchmarked the experimentally derived bounds against classical simulations based on the learned generator and find overall consistency between experiment and theory. Extending such bounds to longer evolution times, and developing protocols that yield tighter or more robust constraints, remains an important challenge. A promising direction could be to incorporate measurements of dynamical susceptibilities that quantify the response of observables to controlled perturbations within the inferred uncertainty model, thereby enabling an experimental characterization of how Hamiltonian–Lindbladian uncertainties propagate to observable-level error bounds.

With additional quantum resources, the learning stage could be upgraded to \emph{optimal} Hamiltonian and Lindbladian estimation at the Heisenberg limit, providing correspondingly tighter certification~\cite{Huang2023,Hu2025}. The framework also offers a natural starting point for a feedback loop in which experimental controls are iteratively refined to steer the realized dynamics closer to a desired target model. Moreover, uncertainty propagation could be made more efficient by measuring suitable susceptibilities directly on the quantum device, enabling on-hardware estimation of tighter error bounds.

The present formulation of BEQS assumes a master-equation description of the dynamics, i.e., a Markovian noise model valid for short bath-correlation times. The framework, however, extends naturally to the opposite limit of long bath-correlation times or slowly drifting experimental parameters. In that setting, rather than learning dissipative jump operators, one infers a probability distribution over (slowly) fluctuating Hamiltonian parameters, yielding an analogous uncertainty model for bounded-error simulation. More generally, the approach could be generalized to learn \emph{colored} quantum noise models jointly with the Hamiltonian and Lindbladian directly from experimental data, enabling characterization of non-Markovian environments within the BEQS framework.

The BEQS framework can also be extended to incorporate error bounds arising from systematic (calibration) deviations between the mean Hamiltonian and an intended target model~\cite{Trivedi2024}. Furthermore, it is complementary to additional learning techniques—such as Hamiltonian locality testing~\cite{Bluhm2025}—which determine whether an unknown $n$-qubit Hamiltonian is $k$-local or $\epsilon$-far from all $k$-local Hamiltonians. Such tools could provide size estimates for any unresolved or residual Hamiltonian terms.

Finally, because BEQS naturally extends to digital quantum simulation via learning effective Hamiltonians of Trotterized circuits~\cite{Pastori2022,Phasecraft2025,Phasecraft22025,GonzlezCuadra2023}, it yields physically transparent diagnostics of Trotter and control errors, as well as a direct characterization of circuit noise through learned Lindbladians; see Appendix~\ref{sec:BeqsDigital}. The underlying concepts will remain relevant in the era of quantum error correction and fault-tolerant quantum simulation.

In the future, it will be interesting to explore whether numerical methods introduced in Ref.~\cite{Mishra2024}—which compute classical bounds on expectation values attainable by noisy quantum circuits—can be adapted to analog quantum computation. It would likewise be valuable to investigate how symmetries and the existence of decoherence-free subspaces can be exploited in the verification of analog quantum simulators, and how the BEQS framework performs in parameter regimes close to a phase transition, where the susceptibility of local observables to small perturbations and state-preparation errors is generically enhanced.

\acknowledgments
W.L. and B.V. thank M. Filippone for useful discussions. T.K., T.O., and B.K. and P.Z, acknowledge funding from the BMW Endowment Fund and from the European Union’s Horizon Europe research and innovation programme under the calls HORIZON-CL4-2022-QUANTUM-02-SGA via Grant Agreement No. 101113690 (PASQuanS2.1) and HORIZON-CL4-2021-DIGITAL-EMERGING-02-10 via Grant Agreement No. 101080085 (QCFD). P.Z. thanks for support by Quantum Science Austria – quantA, an Austrian Excellence Programme funded by the Austrian Science Fund (FWF). T.O. acknowledges support from the German Federal Ministry of Education and Research (BMBF) via the funding programme “Quantum technologies – from basic research to market”, project Grant No. 13N16073 (MUNIQC-Atoms). W.L., B.V., and D.S.F. acknowledge funding from the French “Plan France 2030” research programme HQI (Grant No. ANR-22-PNCQ-0002). W.L. is additionally supported by the “QuanTEdu-France” programme (Grant No. ANR-22-CMAS-0001, France 2030). D.S.F. acknowledges financial support from the Novo Nordisk Foundation (Grant No. NNF20OC0059939, Quantum for Life) and by the European Research Council (ERC) via Grant No. 101163938 (GIFNEQ). L.K.J acknowledges support from the European Union’s Horizon Europe program under the Marie Sklodowska Curie Action Project ETHOQS (Grant no. 101151139), and the support of the IQOQI, Innsbruck of the Austrian academy of sciences, during secondment of the ETHOQS project. M.K.J, F. K, J. F, R. B, and C.F.R acknowledge funding from the European Union's Horizon 2020 research and innovation programme under grant agreement No 101113690 (PASQuanS2.1), and via the Austrian Science Fund through the SFB BeyondC (Grant-DOI 10.55776/F71). P.Z. is a member of Q-SenSe at JILA, a National Science Foundation (NSF) Quantum Leap Challenge Institute focused on quantum sensing and engineering.

{\bf Data availability:} The raw data used for Hamiltonian learning of the 10- and 51-ion Hamiltonians and for measuring the observables of the time-evolved N\'eel state of Fig. \ref{fig:TheoryVsExp51} are openly available at \citep{joshi2025boundederror}.

\appendix

\section{Theory of BEQS}\label{app:theory}

This Appendix first reviews the foundations of Hamiltonian and Lindbladian Learning, including the underlying estimation protocols to obtain the statistical models in Eqs.~\eqref{eq:HDistribution}, or~\eqref{eq:statistical_model}. We then proceed to discuss explicit expressions for error bounds in quantum simulation, which form the basis for the experimental analysis in the main text. We also provide an in depth example illustrating a full theory simulation of the BEQS protocol, demonstrating the consistency and applicability of the method, and in close relation to the $10$ ion experiment described in the main text. Finally, we discuss the extension of the BEQS framework to digital architectures.

\subsection{Hamiltonian and Lindbladian Learning}\label{app:HLLL}

In Hamiltonian and Lindbladian Learning, the quantum simulator is viewed as a "black-box" device that prepares non-equilibrium states of the form
$\rho_t = e^{\mathcal{L} t}\rho_0$. 
The experiment begins by initializing high-fidelity product states 
$\ket{\psi_0} = \ket{\psi_1}\!\otimes\dots\otimes\!\ket{\psi_N}$, 
which are evolved under the (unknown) generator $\mathcal{L}$ for a sequence of evolution times 
$t_0,\dots,t_R$. 
At each time step, expectation values of a set of few-body observables $\{P_k\}$ 
(e.g., Pauli strings) are estimated via randomized measurements~\cite{Elben2022}, yielding the dataset
\begin{equation}
\label{eq:PtoH}
\bigl\{
\langle P_k \rangle_{t_0}, \dots, \langle P_k \rangle_{t_R}
\bigr\}_{k=1}^{K}.
\end{equation}

The dynamics of these expectation values are governed by the adjoint master equation (Ehrenfest theorem),
\begin{equation}
\label{eq:Ehrenfest}
\frac{\mathrm{d}}{\mathrm{d}t}\langle P_k \rangle
= \langle \mathcal{L}^\dagger (P_k) \rangle,
\end{equation}
where $\mathcal{L}^\dagger$ is defined through 
$\mathrm{tr}[ \mathcal{L}(\rho) O ] = \mathrm{tr}[ \rho \, \mathcal{L}^\dagger(O) ]$.
We parametrize the generator as
$\mathcal{L}^\dagger \equiv \mathcal{L}^\dagger(\vec{c})$,
expanding it into a basis of Hamiltonian terms $\{h_i\}$ 
and dissipative jump operators $\{L_i\}$.
Equation~\eqref{eq:Ehrenfest} is then linear in the coefficients $\vec{c}$ and can be written as
\begin{equation}
\label{eq:linearsystem}
\vec{b} = \vec{M} \vec{c},
\end{equation}
where $\vec{M}$ and $\vec{b}$ are constructed directly from the measured few-body correlators in Eq.~\eqref{eq:PtoH}.

Below we summarize two approaches of solving Eq.~\eqref{eq:linearsystem} based on either integral estimators~\cite{Olsacher2025} or time-derivative estimators~\cite{Franca2022}. We emphasize that Hamiltonian and Lindbladian Learning, as described here, builds on modest, minimal-overhead experimental quantum resources. These include high-fidelity preparation of product states and single-atom (or ion) single-shot readout in any spin-basis that are generically available on  programmable analog quantum simulators.

\subsubsection{Integral Method}\label{subsec:10IonIntegral}

The integral method~\cite{Olsacher2025} is based on integrating Eq.~\eqref{eq:Ehrenfest} from $0$ to $T$ $(\sim 1/J_0)$ for each observable $P_k$,
\begin{equation}
\label{eq:EhrenfestIntegral}
\langle P_k \rangle_T - \langle P_k \rangle_0 
= \int_0^T \langle \mathcal{L}^\dagger(P_k) \rangle_t \, \d t.
\end{equation}
Inserting the parametrized generator $\mathcal{L}^\dagger(\vec{c})$ yields a linear constraint as in Eq. (\ref{eq:linearsystem}), $\vec{b}=\vec{M}\vec{c}$, with
\begin{equation}
b_k = \langle P_k \rangle_T - \langle P_k \rangle_0, \quad M_{km} = \int_0^T \langle i[h_m, P_k] \rangle_t \, \d t,
\label{eq:constraints}
\end{equation}
for $m=1,\dots,M$, and
\begin{equation}
    M_{km} = \frac{1}{2} \int_0^{T} \expval*{L_m^\dagger [P_k, L_m] + \mathrm{h.c.}}_{t} \, \mathrm{d}t.
\end{equation}
for $m=M+1,\dots,M'$.

Both $\vec{b}$ and $\vec{M}$ are estimated directly from the experimentally obtained time traces of few-body Pauli observables via the trapezoidal rule. As commutators of few-body Pauli operators remain few-body, all required correlators are contained in the dataset~\eqref{eq:PtoH}. Multiple initial states may be used to increase the rank of the constraint matrix if necessary, and linearity ensures robustness to state-preparation errors.

The learned parameters follow from a least-squares estimate,
\begin{equation}
\label{eq:cost_function}
\vec{c}^{\,*}
= \arg\min_{\vec{c}} \Vert \tilde{\vec{M}}\vec{c} - \tilde{\vec{b}} \Vert_2^2,
\end{equation}
where $\tilde{\vec{M}}$, $\tilde{\vec{b}}$ denote experimentally estimated quantities. A regularization term $\beta \|(\mathds{1} - GG^{\mathrm{T}})\vec{c}\|_2^2$ can be incorporated into the minimization, taking the role of a prior belief in Bayesian inference~\cite{Evans2019}. For instance, a matrix $G$ can be constructed such as to incentivize solutions that adhere to a certain structure of the Hamiltonian, e.g., an algebraic decay of interactions that we utilize in Sec.~\ref{sec:ExpResults}, scale-independent properties, such as translation invariance, or parameters averaging over many small $N$-body terms. This provides a natural route for extending BEQS to regimes where fully resolving every individual higher-order term becomes impractical. Details can be found in Ref.~\cite{Olsacher2025} and Sec.~I of the SM~\cite{Supp}.

\begin{figure*}[t!] 
    \centering
    \includegraphics[width=\linewidth]{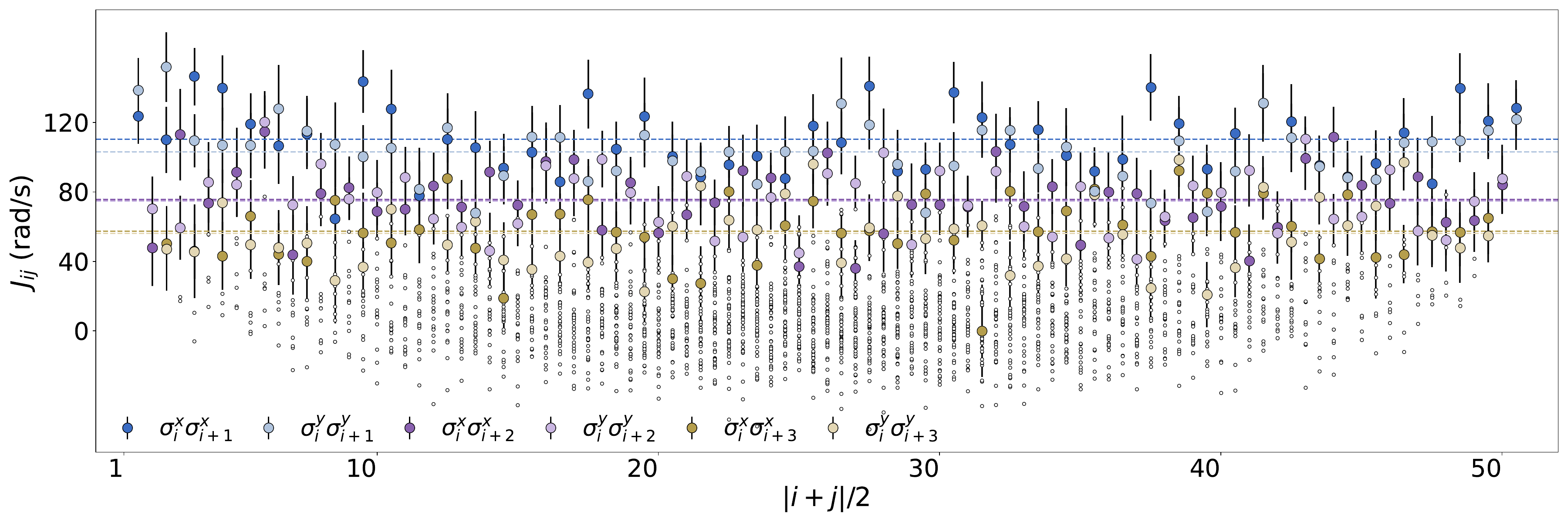} 
    \caption{\textbf{Regularization of Hamiltonian parameters.}
    Here we show the learned coupling parameters $J_{ij}^{(x,x)}$ and $J_{ij}^{(y,y)}$ of the ansatz Hamiltonian in Eq.~\eqref{eq:twobody_Lindbladian}, for $N=51$ qubits, together with their error bar after a regularization has been performed. Here, we have chosen a regularization parameter of $\beta=8$ (see Ref.~\cite{Olsacher2025} and Sec.~I of the Supplemental Material (SM)~\cite{Supp} for details).
    In blue, we show the nearest neighbors (n.n.) terms, in purple, the second n.n. terms, and in yellow the third n.n. terms. Uncolored dots correspond to long range couplings, which we show here with a reduced marker size to enhance clarity and prevent excessive visual density. The dotted lines, indicate the respective mean interactions for fixed distance $d=|i-j|$. We see that the interactions overall decay with the distance $d$, and are relatively homogeneous across the chain, with 18 rad/s average scatter around the mean values. The size of the errorbars are on average $18$ rad/s, compared to $40$ rad/s without regularization. These Hamiltonian parameters, together with the $\sigma_z$ field shown in Fig.~\ref{fig:HL51}(d), is used for the simulations of the error model shown in Fig.~\ref{fig:TheoryVsExp51} in the main text.
   }
    \label{fig:Jijreg51} 
\end{figure*}

\subsubsection{Differential Method}\label{subsec:10IonDiff}

In this section, we show how to access the vector $\mathbf{b}=\{b_k\}_k$, with \mbox{$b_k=\frac{\mathrm{d}}{\mathrm{d}t}\langle P_k \rangle_{\rho_k} \big|_{t=0}$}, defined in Sec.~\ref{subsec:51IonDifferential} with randomized measurements~\cite{Elben2022}. 
Ref.~\cite{Franca2022} presented efficient estimators based on process shadow tomography, in the sense that $\mathbf{b}$, can be accessed with a total measurement budget that is polynomial in system size. Here we use estimators specific to randomized Pauli measurements that allowed us to access $\mathbf{b}$ with reduced statistical errors. 

Our randomized measurement procedure consists of preparing random initial states $\tilde{\rho}_r=\ket{\psi_r}\bra{\psi_r}$, with $\ket{\psi_r}=\bigotimes_l\ket{\psi_{r,l}}$, where $\ket{\psi_{r,l}}$ is a Pauli eigenstate, evolving the system via the Lindbladian for a time $t$ and acquiring $N_M$ bitstrings in random Pauli bases $V_r$. We repeat this procedure for $r=1,\dots,N_U$ to form our dataset of $N_U\times N_M$ bitstrings.

To extract $b_k$, we first 
consider the estimation of $\expval{P_k}$ at given time $t$.
In our dataset, we consider the settings $r\in \mathcal{R}$ such that the measuring setting $V_r$ is compatible with the measurement of $P_k$. 
This means we have access to $\expval{P_k}_r$ which is an unbiased estimator of $\Tr(\tilde{\rho}_r P_k(t))$ up to shot noise errors.
Next, we further restrict our analysis to the settings $\mathcal{R}'\subset \mathcal{R}$ such that the initial state is also ``compatible'' with our mixed $\rho_k$, in the sense that 
$\rho_r^{(s)}=\rho_k^{(s)}$, for $s=l,m$, the qubit pair that $P_k$ is acting on.
In this case, $\expval{P_k}_r$
is an unbiased estimator of $\expval{P_k}=\mathrm{Tr}(\rho_k P_k(t))$, because
\begin{align}
    \mathbb{E}\left[\expval{P_k}_r\right]
    = \mathrm{Tr}
    \left(
    \mathbb{E}_{\mathcal{R}'}(\rho_r)P_k(t)
    \right)
    =\expval{P_k},
\end{align}
where $\mathbb{E}$ denotes the combined ensemble average over the settings $\mathcal{R}'$ and the quantum mechanical average. Moreover, we have used that $\mathbb{E}_{\mathcal{R}'}(\rho_r)=\rho_k$.
Our final estimator for $\expval{P_k}$ is obtained by taking the empirical average of $\expval{P_k}_r$ over all $|\mathcal{R}'|$ settings. In the situation where $|\mathcal{R}'|=0$, the corresponding value of $k$ were not included in the analysis.
Then, after repeating the procedure for all $t$, we interpolate each time trace $\expval{P_k(t)}$ with low degree polynomials to estimate the $b_k$ and construct the vector $\boldsymbol{b}$.

\subsubsection{Regularized $J_{ij}$ couplings for the $51$ ions experiment}

In addition to the learning procedure presented in \ref{subsec:H51}, we further apply Tikhonov regularization (SM~\cite{Supp} Sec.~I~\cite{Supp} and Ref.~\cite{Olsacher2025}) in order to reduce the shot noise and the statistical error bars of the $J_{ij}^{(x,x)}$ and the $J_{ij}^{(y,y)}$ terms. Such regularization is achieved using the model $J_{ij}^{(x,x)}\approx J_{ij}^{(y,y)} \propto e^{-\tau d}$, to incentivize the two body interactions to decay with distance. We show in the Fig~\ref{fig:Jijreg51}, the measured $J_{ij}^{(x,x)}$ and $J_{ij}^{(y,y)}$ coefficients, and their error-bars. Mean values did not change much ($\pm 5$ rad/s), while the error-bars decreased of $20$ rad/s in average. 
The other terms, the one body and the dissipative terms, did not change compared to the unregularized values.

\subsubsection{Reconstructing non-diagonal Lindbladians with Semidefinite Programming}\label{app:SDP}

To ensure the reconstructed Lindbladian yields a cptp evolution, instead of solving the least-squares problem in Eq.~\eqref{eq:cost_function}, we need to solve the following \emph{second-order cone program}
\begin{align}
\vec{c}^{*} = \arg\min_{\vec{c}} \quad & \norm*{ \tilde{M} \vec{c} - \tilde{\vec{b}} }_2^2 \\
\text{s.t.} \quad & \Gamma(\vec{c}) \succeq 0.
\end{align}
Transforming the problem to \emph{epigraph form} and turning the resulting quadratic constraint $\|A c - b\|_2^2 \le t$ into a linear matrix inequality using the Schur complement one obtains the following \emph{semidefinite program}~\cite{BoydVandenberghe2004}
\begin{align}
\vec{c}^{*} = \arg\min_{\vec{c},\, t} \quad &\,t \\[3pt]
\text{s.t.} \quad
&\begin{bmatrix}
t & (\tilde{M}\vec{c} - \tilde{\vec{b}})^{\!\top} \\[3pt]
\tilde{M}\vec{c} - \tilde{\vec{b}} & \id
\end{bmatrix} \succeq 0, \\[8pt]
&\, \Gamma(\vec{c}) \succeq 0.
\end{align}
The above problems can be readily solved using \texttt{CVXPY}~\cite{Agrawal2018} and the solver \texttt{MOSEK}~\cite{mosek}.

\subsubsection{Error models}\label{sec:error_models}

In the main text, we assumed that the uncertainties in the learned dynamical coefficients follow a Gaussian distribution. We now justify this assumption. The coefficients are obtained via the least-squares estimator
\begin{equation}
    \vec{c}^{\,*} = (\vec{M}^{T}\vec{M})^{-1}\vec{M}^{T}\vec{b}.
\end{equation}
When $\vec{M}$ is noiseless---as is the case for the differential method---and $\vec{b}$ is affected by Gaussian noise, the resulting error model is also Gaussian, since $\Delta \vec{c}^{\,*}$ is a linear function of $\Delta \vec{b}$. The vector $\vec{b}$ consists of expectation values estimated from empirical averages over many experimental shots and is therefore well approximated by a Gaussian distribution by the central limit theorem when the number of shots is sufficiently large. We thus conclude that $\Delta \vec{c}^{\,*}$ is well described by a Gaussian distribution whenever $\vec{b}$ is estimated from a sufficiently large data sample.

If $\vec{M}$ is also noisy, as in the integral method, but the perturbation $\Delta \vec{M}$ is small, one can linearize the least-squares estimator $\vec{c}^{\,*}$ for small deviations. In this regime, one finds
\begin{equation}
\begin{aligned}
    \Delta \vec{c}^{\,*}
    &= (\vec{M}^{T}\vec{M})^{-1}\vec{M}^{T}
       \bigl(\Delta \vec{b} - \Delta \vec{M}\,\vec{c}\bigr) \\
    &\quad
       + (\vec{M}^{T}\vec{M})^{-1}\Delta \vec{M}^{T}
       \bigl(\vec{b} - \vec{M}\vec{c}\bigr).
\end{aligned}
\end{equation}
as shown in Ref.~\cite{Bjrck2024}. This expression makes explicit how deviations from Gaussianity in $\Delta \vec{c}^{\,*}$ arise from the terms involving $\Delta \vec{M}$. The magnitude of these contributions can be bounded by estimating the size of the entries of $\Delta \vec{M}$, which, for a given shot budget, can be controlled with high probability using standard concentration inequalities. 

We emphasize that, due to the substantial amount of post-processing involved, parametrizing the error model for $\Delta\vec{M}$, $\Delta\vec{b}$, and ultimately $\Delta\vec{c}^{\,*}$ in terms of simple Gaussian random variables is intractable. For this reason, we employ nonparametric techniques such as bootstrapping and jackknife resampling. Moreover, in situations where the number of shots is not large enough to invoke the central limit theorem or to ensure that $\Delta \vec{M}$ is negligible, we can still assess Gaussianity directly by performing statistical tests on the bootstrapped and/or jackknifed samples, for instance using the Shapiro–Wilk test.

\subsection{Bounded-error quantum simulation}\label{sec:boundedError}

A central objective of this work is to derive \emph{rigorous and efficiently computable} bounds on the errors that arise when estimating $\mathcal{L} \mapsto \langle O_t \rangle$, given that the implemented dynamics is known only through a learned ensemble ${\cal L}(\vec{g})$ [Eq.~\eqref{eq:statistical_model}]. We provide bounds on (i) the \emph{expected simulation error} and (ii) the \emph{fluctuations} around this mean [Eqs.~\eqref{eq:ExpectedError}--\eqref{eq:TailBound}], without requiring classical sampling over the ensemble. Short-time bounds are derived via a second-order expansion, while long-time bounds depend only on the evolution time and the learned parameter covariance.

To keep the following discussion concise we consider evolution under an ensemble $H'(\vec{g})$, but emphasize that all bounds can be directly extended to dissipative dynamics. First, we consider short-time regimes in which the dynamics can be described by a second-order expansion of $e^{-iH'(\vec{g})t}$. The resulting bounds explicitly depend on the initial state, $\rho_0$, but can be efficiently computed whenever efficient tomography of the initial state is possible. Then, we discuss more general bounds that only depend on the evolution time $t$ and the learned covariance matrix $\Sigma$ of the Hamiltonian and Lindbladian ensemble.

\subsubsection{Bounds for short-time evolution} \label{sec:short}

Consider a learned Hamiltonian ensemble $H'(\vec{g})$ and assume $g_{i}$ to be independent sub-Gaussian random variables with uniform variance proxy $\delta^{2}$. Recall, a random variable $g$ is called sub-Gaussian if there exists a positive constant $C$ such that $\operatorname{Pr}(|g| \geq t) \leq 2 \exp{(-t^2/C^2)}$ for all $t\geq 0$. In Sec.~III of the SM~\cite{Supp} we derive a short-time expansion of $\expval{O'_{t} (\vec{g})}-\mathds{E}[\expval{O'_{t}(\vec{g})}]$ up to second order in time. We then bound linear terms using sub-Gaussian concentration inequalities, and quadratic terms using the Hanson-Wright inequality~\cite{Rudelson2013}, which states a concentration bound for quadratic forms in sub-Gaussian random variables. We obtain that with probability at least $1-\eta$
\begin{equation}\label{eq:HWAppendix}
\bigl|\mathds{E}\left[\expval{O'_{t}(\vec{g})}\right]-\expval{O'_{t}(\vec{g})}\bigr| 
\leq t \delta \norm{\vec{b}_t}_2 \sqrt{L} + \frac{t^2}{2}\delta^2\mu^{\rm HW}_{\eta,D}.
\end{equation}
with
\begin{equation}
    \mu^{\rm HW}_{\eta,D} = \max\qty{\norm{D}_{\rm HS}\sqrt{\frac{1}{\kappa}\log{\frac{4}{\eta}}}, \norm{D}_{\rm op}\frac{1}{\kappa}\log{\frac{4}{\eta}}},
\end{equation}
$L=\frac{1}{c}\log{\frac{4}{\eta}}$, and $\vec{b}_{t}=\vec{b}-\frac{t}{2} \vec{s}$. Here, $\norm{D}_{\rm HS}=\sqrt{\Tr D^{T} D}$ 
denotes the Hilbert-Schmidt norm, and $\norm{D}_{\rm op}=\lambda_{\rm max}(D)$ denotes the spectral norm, i.e., the largest singular value. The constants $c,\kappa \sim \mathcal{O}(1)$ are universal, though their precise numerical value varies across the literature and affects only the tightness of the exponential tail. In practice, $c$ and $\kappa$ may be mildly calibrated to achieve a sharper fit to empirical data. For the Gaussian case $c=1/2$ due to the Chernoff bound. For the Hanson-Wright inequality a recent paper proves a value of at least $\kappa=0.145$~\cite{Moshksar2025}, and a textbook provides $\kappa=1/8$~\cite{Giraud2021}.

Let us briefly describe what is needed to compute the above bound. The vectors $\vec{b}$ and $\vec{s}$ and the matrix $D$ are defined though expectation values of few-body observables evaluated in the initial state. Specifically, one finds that
\begin{equation}
b_{i}=i \Tr(\rho_{0}\left[V_{i}, O\right]), \quad s_{j}=\operatorname{Tr}\left(\rho_{0}\left[\left[H, V_{j}\right], O\right]\right),
\end{equation}
and
\begin{equation}
    D_{i j}(O)=\operatorname{Tr}\left(\rho_{0}\left[V_{i},\left[V_{j}, O\right]\right]\right).
\end{equation}

The bounds derived above can be directly generalized to error bounds on the fidelity 
\begin{equation}
  F(t)=\abs{\braket{\psi_t}{\psi'_t}}^2  
\end{equation}
between the expected wavefunction $H_{\rm learned}\mapsto\ket{\psi_t}$, and a wavefunction from the ensemble $H'(\vec{g})\mapsto\ket{\psi'_t}$. Using again a second-order expansion and the Hanson-Wright inequality one can show that with probability at least $1-\eta$
\begin{eqnarray}
     1-t^2\delta^2[\Tr(C)+\mu^{\rm HW}_{\eta,C}] &\leq &F(t) \\  & \leq& 1-t^2\delta^2[\Tr(C)-\mu^{\rm HW}_{\eta,C}], \nonumber
\end{eqnarray}
where $C_ {ij}={\rm Re}\bra{\psi_0}V_iV_j\ket{\psi_0}$. For a proof, see Sec.~IV of the SM~\cite{Supp}.

\subsubsection{Bounds for long-time evolution}\label{sec:long}

In a recent work, Cai, Tong, and Preskill~\cite{Cai2024} analyzed a physical error model in which the implemented Hamiltonian deviates from the target Hamiltonian in a random but unbiased manner. Their model takes the form of Eq.~\eqref{eq:HDistribution}, which in our context is interpreted as the learned uncertainty in the Hamiltonian. In our setting, the resulting bounds can be reinterpreted as providing guarantees on the expected simulation error and on the concentration of deviations [Eqs.~\eqref{eq:ExpectedError} and~\eqref{eq:TailBound}] arising from uncertainty in the Hamiltonian parameters.
 
Let us briefly review some of the results of Ref.~\cite{Cai2024} and explain how they can be modified to yield bounds for BEQS. Ref.~\cite{Cai2024} assumes purely Hamiltonian evolution, and very mild assumptions on the random variables $g_i$. Here, we will be interested in Gaussian variables, but the results also apply to deformed Gaussian error models. In the case of independent Gaussian variables $g_i$ with uniform variance $\delta^2$ the expected error is bounded by
\begin{equation}\label{eq:bound_expected_app}
    \bigl|\expval{O_t}-\mathds{E}[\expval{O'_t(\vec{g})}]\bigr|\leq [e^{{2t^2\delta^2 M}}-1]\norm{O},
\end{equation}
whereas the concentration around the expected value $\mathds{E}[\expval{O'_t(\vec{g})}]$ is bounded by
\begin{equation}\label{eq:bound_deviation_app}
    \bigl|\mathds{E}[\expval{O_t'(\vec{g})}]-\expval{O_t'}\bigr|\leq 2t\delta\sqrt{M}c\norm{O},
\end{equation}
with probability at least $p=1-2e^{-c^2/2}$. Ref.~\cite{Cai2024} furthermore shows, using Lieb-Robinson bounds, that for geometrically local Hamiltonians these bounds become system-size independent, a highly desirable property for quantum simulations approaching the thermodynamic limit. Specifically, in the above bounds one can replace $M$ by the effective size of the observables past light-cone, and include an appropriate truncation error, that is obtained from Lieb-Robinson bounds. Moreover, similar bounds are obtained for the fidelity between the expected wavefunction, and the ensemble wavefunctions.

In Hamiltonian and Lindbladian Learning several extensions need to be considered. In case the variances vary over the $g_i$ one can replace $M\delta^2\mapsto\sum_{i=1}^M \delta_i^2$ (see Ref.~\cite{Cai2024}, Eq.~(12)). Due to the regularization correlations between the reconstructed parameters can occur, albeit here rather small. For large correlations, i.e., non-diagonal covariance, the above bounds still hold with $M\delta^2\mapsto\norm*{\Sigma^{\frac{1}{2}}}_{L_{1,2}}$, where $\norm*{A}_{L_{p,q}}^q=\sum_{j=1}^n \left( \sum_{i=1}^m |A_{ij}|^p \right)^{\frac{q}{p}}$ (see  Sec.~II of the SM~\cite{Supp}). Moreover, the Lindbladian can be included in these bounds by replacing $M\rightarrow M'$, i.e. including the parameters of the Lindbladian.

\subsubsection{Bounds for ensembles including biases}\label{app:systematicErrorBound}
Let us now consider more general ensembles including a bias $\vec{k}$, i.e., where $\vec{g} \sim \mathcal{N}(\vec{k}, \Sigma)$ (see discussion below Eq.~\eqref{eq:HDistribution} in the main text). These error lead to additional systematic errors in observable expectation values \emph{beyond} the expected error due to parameter uncertainty in Eq.~\eqref{eq:ExpectedError}. In the following we want to bound the error between $\expval{O_t}$ predicted by the mean Hamiltonian $H_{\rm learned}$, and $\expval{O'_t(\vec{g})}$ predicted by ensemble Hamiltonians $H'=H_{\rm learned}+V(\vec{g})$, where now $\vec{g}\sim\mathcal{N}(\vec{k},\Sigma)$. We can split the total error into an error due to the shifted mean, and the parameter uncertainty; we have
\begin{equation}
    \abs{\expval{O_t}-\expval{O'_t(\vec{g})}}\leq \abs{\expval{O_t}-\expval{O''_t(\vec{k})}} + \abs{\expval{O''_t(\vec{k})}-\expval{O'_t(\vec{g}})},
\end{equation}
where $\expval{O''_t(\vec{k})}$ is the expectation value predicted by $H''=H_{\rm learned}+V(\vec{k})$. The first term represents errors due to shifted parameters with respect to $H_{\rm learned}$, while the second terms represents errors due to parameter uncertainties, which can be bounded using the bounds explained above. The first term can be bounded as follows. We have that $\abs{\expval{O_t}-\expval{O''_t(\vec{k})}}\leq\norm{\rho}_{\rm tr}\norm{O_t-O''_t(\vec{k})}$, with $\norm{\rho}_{\rm tr}=1$. Duhamel's formula~\cite{Cai2024} states that
\begin{equation}
    O_t-O''_t(\vec{k}) = - i\int_0^t \d s \;e^{iH''(t-s)} \, [V(\vec{k}), O_s] \, e^{-iH''(t-s)}.
\end{equation}
Then, using the fact that $\norm{e^{iXt}}=1$, and $\norm{[X,Y]}\leq 2\norm{X}\norm{Y}$ one can show that
\begin{equation}
    \abs{\expval{O_t}-\expval{O'_t(\vec{k})}}\leq 2t\norm{k}_1\norm{O}.
\end{equation}

\subsection{Detection of Systematic Errors}\label{app:systematic_errors}

\begin{figure*}
    \centering
    \includegraphics[width=1\linewidth]{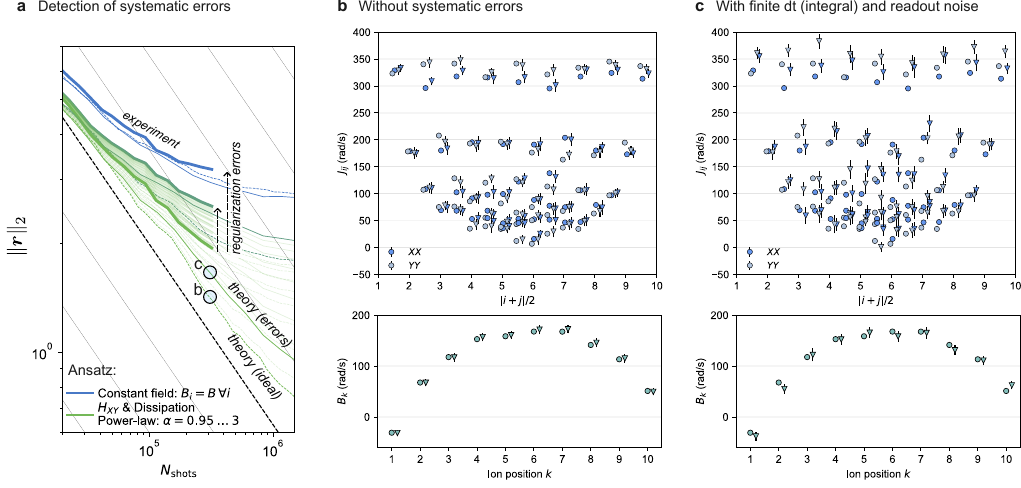}
    \caption{{\bf Detection of systematic errors.} ({\bf a}) Here, we demonstrate in more detail the possibility of detecting systematic errors by observing the residual norm, $\norm{\vec{r}}_2$, as a function of the number of experimental shots, $N_{\rm shots}$. We begin with the green lines representing the ``optimal'' ansatz obtained in the main text and Fig.~\ref{fig:Hlearning10}. Again, the bold solid line represents experimental data, the dashed line represents an ideal resimulation of the experiment with negligible integration error and without measurement errors, and the thin solid line represents a resimulation with a finite step size $\d t$ and a simulated $2\%$ readout error, beginning to affect the scaling at roughly $10^5$ shots. Moreover, we can detect biases arising from inadequate regularizations. For instance, choosing an excessively large power-law decay—here up to $\alpha = 3$—induces a deviation from the ideal scaling and ultimately leads to plateaus in the residual norm. We also consider a regularization enforcing a site-independent magnetic field (blue lines), with $B_i = B$ for all $i = 1, \dots, 10$, which exhibits an even more pronounced deviation from the ideal scaling.
    To study the effect of systematic errors in the learning procedure, we use the experimentally learned Hamiltonian and Lindbladian as a model system and resimulate the learning procedure, once without systematic errors and once including them. In ({\bf b}), we show the reconstructed Hamiltonian from a simulation with twice the experimental time resolution and no measurement or readout errors. In this case, the learned Hamiltonian (triangles) is in perfect agreement with the simulated model (circles) within the error bars. In ({\bf c}), we show the reconstructed Hamiltonian from a simulation with the same time resolution as in the experiment and including a simulated $2\%$ readout error. In this case as well, the majority of parameters agree with the model parameters within the error bars; only the nearest-neighbor couplings exhibit a slight bias. In the real experiment, we expect this bias to be smaller due to much smaller readout noise and the fact that the obtained error bounds in Fig.~\ref{fig:BoundedError} already capture the variation of the experimental data very well.
    }
    \label{fig:SystematicErrors}
\end{figure*}

In our analysis, we have assumed that systematic errors are negligible compared to the residual parameter uncertainty arising from the finite number of experimental shots. This leads to the error model in Eq.~\eqref{eq:statistical_model} in the main text, where the uncertainty is characterized by $V(\vec{g})$ with $\vec{g}\sim N(0,\Sigma)$. In the following, we justify this assumption and outline in more detail how to diagnose and detect systematic errors that are present in any real experiment.

Various systematic errors may arise in Hamiltonian learning; for instance, (i) an inadequate operator ansatz for the Hamiltonian or Lindbladian, (ii) inadequate regularization of the learning equation, or (iii) measurement errors and readout noise (see also Appendix~\ref{app:ExpImperfections}). In the main text, specifically in Fig.~\ref{fig:Hlearning10}, we have shown that choosing an inadequate operator ansatz can be detected by considering the residual norm (or \emph{learning error}), $\norm{\vec{r}}_2$, as a function of the number of experimental shots, $N_{\rm shots}$. If missing ansatz operators are small and all relevant terms are contained in the ansatz, the residual norm scales as $1/\sqrt{N_{\rm shots}}$~\cite{Olsacher2025}. Indeed, choosing an inadequate ansatz leads to plateaus, as seen in Fig.~\ref{fig:Hlearning10}.

The scaling of the residual norm can also be used to detect inadequate regularization of the learning equations. To see this, we consider the ``optimal'' ansatz obtained in Sec.~\ref{subsec:ExpResultsIntegral} and shown in Fig.~\ref{fig:Hlearning10}: the XY Hamiltonian with an additional site-dependent field, and a Lindbladian with collective and local dephasing as well as spontaneous decay (green lines). We can induce an inadequate parameterization of the learning equation by increasing the power-law decay parameter $\alpha$ beyond its optimal value of $\alpha=0.95$ determined in the main text. This effectively attempts to fit an excessively strong power-law decay, $J_{ij}\propto \abs{i-j}^{\alpha}$, leading to a deviation from the ideal shot-noise scaling, and eventually resulting in a plateau, see Fig.~\ref{fig:SystematicErrors}(a). Similarly, one can choose an ansatz with a site-independent magnetic field, i.e., $B_i=B$ for all $i=1,\dots,10$, which results in the blue curves in Fig.~\ref{fig:SystematicErrors}(a) and leads to an even stronger deviation.

Moreover, the scaling can be used to detect measurement errors and readout noise~\footnote{We emphasize here, that not all measurement errors can necessarily be detected. For instance, it may happen, that certain measurement errors alter the structure of the inferred Hamiltonian but nevertheless lead to an observed dynamics that can be described by a Hamiltonian and Lindbladian, even in the limit of infinitely many experimental shots.}. To see this, we compare two resimulations of the learning procedure. As a model system we choose the Hamiltonian learned from experimental data in Sec.~\ref{subsec:ExpResultsIntegral}, which takes the form 
\begin{equation}\label{eq:XYwithB}
H_{XY} = \sum_{i<j} J_{ij}\,(\sigma_i^x \sigma_j^x + \sigma_i^y \sigma_j^y) + \sum_{i} B_i \sigma_i^z,
\end{equation}
with parameters \(J_{ij}\) and \(B_i\) as shown in Fig.~\ref{fig:Hlearning10}(b) and (c) respectively. Moreover, we include the dominant dissipation as collective dephasing with $\Gamma_{\rm col} \approx 0.15\,J_0$ as shown in Fig.~\ref{fig:Hlearning10}(d). In the first simulation, we choose twice the experimental time resolution and no measurement errors. This yields an ideal shot-noise scaling, see thin green solid line in Fig.~\ref{fig:SystematicErrors}(a). In the second simulation, we choose the same time resolution as in the experiment and include a $2\%$ readout noise, as explained in Sec.~\ref{subsec:ExpResultsIntegral} in the main text. This error can also be detected as it leads to a deviation from the ideal shot-noise scaling above roughly $10^5$ shots, which closely resembles the scaling observed in the experimental data (see the green solid and bold solid lines respectively).

Finally, to asses the effect of systematic errors in the learned coefficients we can compare the reconstructed parameters that we obtain from both simulations at the same measurement budget of $3.2\times 10^5$ shots, and compare them against the model parameters. In Fig.~\ref{fig:SystematicErrors}(b) we show the parameters obtained from the ideal simulation without systematic errors. Here, the parameters are in perfect agreement with the model parameters within the error bars. In the second simulation the majority of reconstructed parameters agree with the model parameters within the error bars, except nearest-neighbor terms that exhibit a slight bias toward larger values. We emphasize that in the real experiment these biases are likely smaller as the readout noise is much smaller than $2\%$ and the simulated error bounds in Fig.~\ref{fig:BoundedError} well agree with the variation of the experimental data.

\subsection{Classical Simulation of BEQS protocol}

\begin{figure*}[t]
    \centering
    \includegraphics[width=1\linewidth]{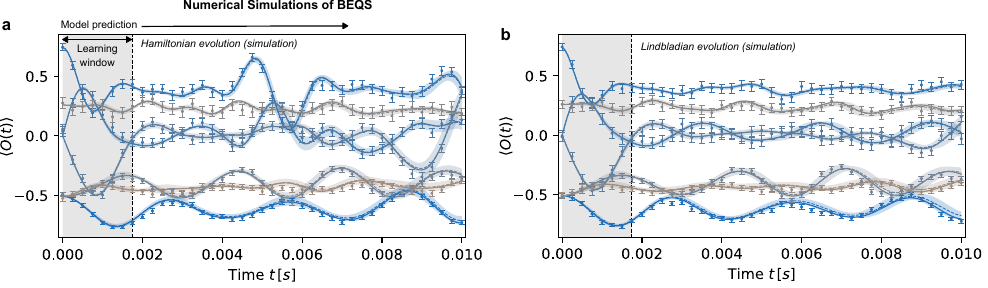}
    \caption{Illustrations of BEQS for systems of $N=10$ spins. Numerical simulations of complete BEQS experiments for (a) Hamiltonian evolution and (b) Lindbladian evolution. The Hamiltonian and Lindbladian parameters correspond to the (mean) parameters of the experimentally learned Hamiltonians and Lindbladians in Fig.~\ref{fig:Hlearning10} in the main text. In the learning mode, these Hamiltonians and Lindbladians are learned back from simulated data using a measurement budget of $3.2\cdot 10^5$ shots located in the learning window (grey area) using the method described in Appendix~\ref{subsec:10IonIntegral}. The learning parameters, i.e., initial states, measurements and temporal resolution of measurements, is close to the experimental protocol (see Appendix~\ref{app:IntegralMethodExp} and main text). In the computational mode we extend the simulated data beyond the learning window to compare against the simulation of the learned Hamiltonian and Lindbladian ensembles $H'(\vec{g})$, and ${\cal L}'(\vec{g})$ respectively. We show here a selection of one- and two-qubit correlation functions, where markers represent the simulated measurements, solid lines correspond to the trajectories predicted by the learned Hamiltonians and Lindbladians, dashed lines denote ensemble averages, and shaded regions represent $90\%$ prediction intervals estimated from $N_s=50$ samples.}
    \label{fig:beqs_resim}
\end{figure*}

To make the concept and internal consistency of BEQS more transparent, we illustrate the method using a system of \(N = 10\) spins. For this system size, the complete two-step protocol can be simulated numerically, enabling also a direct comparison with experimental data. Figures~\ref{fig:beqs_resim}(a) and~(b) show the results for the Hamiltonian and Lindbladian settings, respectively.

For our simulation, we take the mean Hamiltonian learned from the experiments in Sec.~\ref{subsec:ExpResultsIntegral}, which is specified in Eq.~\eqref{eq:XYwithB}. The Lindbladian model is chosen analogously, corresponding to collective dephasing with rate \(\Gamma_{\rm col} \approx 0.15\,J_0\) [see Fig.~\ref{fig:Hlearning10}(d)]. Using the experimentally learned Hamiltonian and Lindbladian as our model ensures that the illustrative simulations remain directly comparable with the experiments discussed in Sec.~\ref{sec:ExpResults} in the main text.

Using this model, we simulate a complete BEQS experiment, with the numerical simulation serving as a digital clone of the quantum simulator. The first stage corresponds to the Hamiltonian (and Lindbladian) \emph{learning mode}, in which the statistical quantum model is inferred from simulated experimental data over a finite time window (indicated by the gray region in Fig.~\ref{fig:beqs_resim}). In the second stage, the simulator is operated in \emph{computational mode} to estimate observables \(\langle O \rangle\), which are then compared with the predictions of the learned statistical model (white region in Fig.~\ref{fig:beqs_resim}).

In the learning mode, we adhere to the experimental protocol (see Appendix~\ref{app:IntegralMethodExp}) and prepare two distinct product states, which are subsequently evolved under the model Hamiltonian and Lindbladian at discrete times \(t_k \in \{0, 0.25, \dots, 1.5\}\,\mathrm{ms}\). At each time step, randomized measurements are performed to estimate all one-, two-, and three-qubit Pauli correlation functions (representative data points within the learning window are indicated in Figs.~\ref{fig:beqs_resim}a and~b). The total measurement budget is fixed to \(3.2\times10^5\) shots to match the experimental conditions (see Appendix~\ref{app:IntegralMethodExp} for details)\footnote{In the numerical simulations, twice the number of integration points is used to suppress numerical integration errors, while the number of shots per point is reduced accordingly to keep the total measurement budget unchanged.}. The resulting data are processed using the procedure described in Sec.~\ref{subsec:10IonIntegral} to infer the learned ensembles \(H'(\vec{g})\) and \({\cal L}'(\vec{g})\). The covariance matrix $\Sigma$ is obtained via $N_b=50$ bootstrap resamples, as explained in Appendix~\ref{sec:error_models}.

In the computational mode, we use the quantum simulator to compute observable expectation values $\expval{O}$. In both simulations, the one with and without Lindbladian, we extended the randomized measurement procedure beyond the learning window for up to $10$ ms, giving us access to the determination of all up to three-body correlation functions \footnote{Here, as in the experiment, these measurements were performed without taking any symmetry of the dynamics into account.  Note that computing (measuring) fewer observables would require less shots.}. This required an additional measurement budget of $1.3\times 10^6$ shots. We compare some of the computed expectation values with the expectation values predicted by our learned models. To this end, we simulate the learned ensembles $H'(\vec{g})$ and ${\cal L}'(\vec{g})$, as task which can still be performed classically. Each simulation shows the predicted trajectories under the (mean) learned Hamiltonian and Lindbladian (solid lines), as well as the average trajectory (dashed lines). For some observables we can observe a difference between these trajectories which results in the expected error in Eq.~\eqref{eq:ExpectedError}. The concentration of errors is indicated by the $90\%$ prediction intervals (shaded areas), which remains well controlled for even up to $10$ms.

We observe agreement between the computed expectation values and model predictions, both in theory simulations as well as in the experiment (see Fig.~\ref{fig:BoundedError}c in the main text), demonstrating the internal consistency of the BEQS framework.

\subsection{BEQS in Digital Architectures}\label{sec:BeqsDigital}

BEQS as developed above, extends naturally to digital quantum simulation, both with and without error correction. In digital quantum simulation, the Hamiltonian of interest, $H = \sum_k H_k$, is implemented through a sequence of quantum gates organized into Trotter blocks. For example, one may define
\begin{equation}\label{eq:Trotter}
    U_\tau = \prod_k e^{-i H_k \tau} \equiv e^{-i H_F(\tau) \tau}.
\end{equation}
As $\lim_{n \to \infty} [U_{t/n}]^n = e^{-i H t}$ holds, repeated application of the Trotter block, $ U_\tau$ in the limit $\tau \to 0$, reproduces the target evolution exactly.
Trotterization thus assumes the structural form of a Floquet problem, where $H_F(\tau)$ in Eq.~(\ref{eq:Trotter}) represents the effective Hamiltonian realized in the simulation for a finite Trotter step. The Magnus expansion,
$H_F(\tau) = \sum_{l} \Omega_l \tau^l$,
coincides at leading order with the target Hamiltonian, $\Omega_0 = H$, while higher-order terms, $\Omega_{l>0}$, are responsible for the intrinsic Trotter errors.

Learning the operator structure of the Floquet Hamiltonian (or, in extended form, the Lindbladian)~\cite{Pastori2022} thus provides direct physical insight into both calibration and Trotter errors, facilitating the debugging and characterization of digital quantum simulation circuits. This approach contrasts with global quantities, such as process fidelities, which offer only limited knowledge about the effective Hamiltonian. Notably, our framework remains applicable in the era of error correction and fault-tolerant digital quantum simulation.

\section{Experimental platform}\label{app:exp}

The trapped-ion analog quantum simulator used for Hamiltonian learning uses a linear radio-frequency trap for confining linear strings of singly-charged $^{40}$Ca ions in an anisotropic harmonic potential with oscillations frequencies of $\omega_x =2\pi\times 2.917$~MHz and $\omega_y =2\pi\times 2.886$~MHz in the transverse directions. The longitudinal oscillation frequency was set to $\omega_z =2\pi\times 0.217$ MHz for strings of $N=10$ ions and to $\omega_z =2\pi\times 0.115$ MHz for experiments with $N=51$ ions, resulting in ion strings with lengths of $71\,\mu$m and $263\,\mu$m, respectively. The ion crystals are cooled to low temperature by Doppler cooling, followed by polarization-gradient cooling of the longitudinal modes and sideband cooling that prepares all transverse modes close to their ground state \cite{Kranzl2022}. 

\subsection{Engineered many-body interactions}
Qubit states $|{1}\rangle$ and $|{0}\rangle$ are encoded in the two $| \texttt{S}_{1/2} ,m=+1/2\rangle$ and  $| \texttt{D}_{5/2} ,m=+5/2\rangle$ Zeeman sublevels that are linked by a quadrupole transition with a wavelength of 729~nm. Local qubit control is achieved by a strongly focused steerable beam off-resonantly coupling the states that induces a differential light shift on a selected ion. 

A long-range Ising interaction is mediated by a bichromatic beam of a narrow-frequency laser that illuminates the ions from a direction perpendicular to the ion string with a Gaussian beam waist (radius) of $w_z=105\mu$m for a 10-ion chain ($w_z=310\mu$m for a 51-ion chain) along the ion string and couples all qubits off-resonantly to all transverse motional modes. The two frequency components of the laser are detuned from the red and blue $x$ center-of-mass mode sidebands with $-\Delta+\delta_c$ and $\Delta+\delta_c$, respectively. The interaction can be described by an effective Ising interaction \cite{Porras2004,Britton2012}, 
\begin{equation}
    H_{\rm Ising}=\sum_{i<j} J_{ij} \sigma^x_i \sigma^x_j,
    \label{eq:H_Ising}
\end{equation}
with
\begin{equation}\label{eq:JijTheory}
    J_{ij} = \Omega_i\Omega_j\frac{\hbar k^2}{2m}\sum_{l=1}^{2N}
    \frac{b_{i,l}b_{j,l}}{(\Delta+\max(\omega_l))^2-\omega_l^2},
\end{equation}
ion-dependent carrier Rabi frequencies \textbf{$\Omega_i$}, wave number $k$, ion mass $m$, normal mode frequencies $\omega_l$, and $\Delta$, which denotes the detuning of the laser tones with respect to the oscillation frequency of the center-of-mass mode in the $x$-direction. 
The mode participation vectors $b_{i,m}$ are normalized to the cosine of the angle between the directions of oscillation and the laser beam propagation.
The coupling matrix can be approximated by a power-law interaction, $J_{ij}=J_0/{|i-j|^\alpha}$ \cite{Britton2012}.
The Pauli spin matrices appearing in eq.~(\ref{eq:H_Ising}) can be replaced by $\sigma^\phi=\cos\phi\,\sigma^x+\sin\phi\,\sigma^y$ by shifting the phase of the laser by an amount $\phi$.
For a peak Rabi frequency of $2\pi\times 115$~ kHz and a detuning of $\Delta= 2\pi\times 25$~kHz, we find $J_0=576$~rad/s and $\alpha = 1.19$ for a 10 ion-chain. Similarly, for a 51 ion chain, with a peak Rabi frequency of $2\pi\times 77$~ kHz and a detuning of $\Delta= 2\pi\times 25$~kHz, we find $J_0=185$~rad/s and $\alpha = 0.97$. As the light fields mediating the interaction also off-resonantly couple to dipole transitions, they induce spatially varying light shifts that adversely affect the desired spin-spin interaction. Those light shifts are compensated by a third tone of the laser, which is detuned by $\omega_{comp}=2\pi\times 1.4$~MHz from the carrier transition, that induces light shifts of equal magnitude but opposite sign by off-resonant excitation of the carrier transition. The beam, however, also off-resonantly couples to first-order sideband transitions of the transverse motional modes, giving rise to $(a_m^\dagger a_m+\mathcal{I})\sigma^z$ terms where $a_m$ denotes an annihilation operator of phonons in the transverse mode $m$. In the case where low-frequency transverse modes have frequencies approaching $\omega_{comp}$, these shifts can become noticeable.

If the center frequency of the bichromatic laser is shifted by an amount $\delta_c\gg J_0$, the transitions $\ket{0}_i\ket{0}_j\leftrightarrow \ket{1}_i\ket{1}_j$ are suppressed so that the Ising interaction (\ref{eq:H_Ising}) turns into a flip-flop interaction \cite{Jurcevic2014},
\begin{equation}
    H_{XY}=\sum_{i>j} J_{ij} (\sigma^+_i \sigma^-_j+\sigma^-_i \sigma^+_j)=\frac{1}{2}\sum_{i>j} J_{ij} (\sigma^x_i \sigma^x_j+\sigma^y_i \sigma^y_j).
    \label{eq:H_XY}
\end{equation}

Additionally, first-order Zeeman and electric quadrupole shifts can give rise to spatially varying field terms,
\begin{equation}
    H_Z =  \sum_i B_i \sigma^z_i,
\end{equation}
which will have to be added to the Hamiltonians (\ref{eq:H_Ising}) and (\ref{eq:H_XY}).
Further details about the experimental setup can be found in references \cite{Kranzl2022,Joshi2022}.

\subsection{Local control}
A strongly focused steerable laser beam with a beam diameter much smaller than the ion-ion distance is employed for single-qubit $\sigma^z$ operations by light-shifting a qubit's transition frequency. In combination with collective spin rotations, it enables the preparation of the ions in arbitrary input product states.  

The quantum state of the qubits is measured with high fidelity in the $z$-basis by spatially resolved measurements of the ions' fluorescence. Measurements of arbitrary N-qubit Pauli strings can be accomplished by locally rotating the qubits prior to the fluorescence read-out.

\subsection{Implementation of the integral method}\label{app:IntegralMethodExp}

For the Hamiltonian learning of a ten-ion system with the integral method, the input state is prepared with a global laser beam, performing single-qubit rotations on each qubit simultaneously. The input state is represented by $\ket{\psi_0}=e^{-i/2\sum j\theta_j (\hat{n}\cdot\sigma_j)}  \ket{1} ^{\otimes N}$, where 
$\theta$ is the angle of rotation, and the unit vector $\hat{n}$ denotes axis of rotation. For Hamiltonian learning, the two input states $(\theta = 0.33 \pi, \hat{n}=-\mathbf{e}_y)$ and $(\theta = 0.5 \pi, \hat{n}=\mathbf{e}_x)$ are chosen 

Each state is time-evolved under the engineered interaction for $t\in\{0,0.5,1,1.5\}$~ms and  
measured in a random Pauli basis $P^\alpha = \bigotimes_{i=1}^N P_i^\alpha, \quad P_i^\alpha \in \{\sigma_i^x, \sigma_i^y, \sigma_i^z\}$.
For each time-evolved state, a total of measurement settings $N_u=$~200 was used and each measurement repeated $N_m=200$ times. In post processing, up to three-qubit observable expectation values are estimated from raw data to evaluate integrals and their expectation value differences to solve Lindblad equation of motion. A total of $3N$ single-qubit observables and $9N(N-1)/2$ two-qubit observables are extracted from the raw measurements. Depending upon the chosen ansatz, we estimate integrals for only corresponding commutators that are necessary and carry out Hamiltonian learning with those $O(t)-O(0)$ for which a complete set of commutators are available in the measurement settings. 

\subsection{Implementation of the differential method}
For the Hamiltonian learning of a $N=51$-ion system with the differential method, input states $\ket{\psi_0^\alpha}= \prod_j U_j^\alpha  \ket{1} ^{\otimes N}$, $\alpha=1\ldots M$, are prepared where unitaries $U_j^\alpha$ are randomly chosen to prepare ion $j$ in any of the Pauli eigenstates with equal likelihood. Next, the states $\ket{\psi_0^m}$ are time-evolved for various durations (0 to 1 ms with a total of $N_t=$ 11 time steps) and measured in random Pauli bases. 
In total, $N_u=200$ different combinations $\{\ket{\psi_0^\alpha},P^\alpha\}$ are prepared and measured for $N_m=200$ experimental repetitions. From each experimental setting, observable expectation values for a reduced state is then used to evaluate up to two-qubit observable expectation values. Furthermore, a time trace for each observable is fitted to a degree 2 polynomial function ($O(t)=a_0 + a_1t+a_2t^2$) and coefficient $a_1$ (slope) is estimated from the fit. 

\subsection{Experimental imperfections}\label{app:ExpImperfections}
Collective qubit rotations by resonant laser pulses are impacted by the inhomogeneous intensity profile of the laser beam, which results in over- and under-rotations of certain qubits. For the 10-ion experiments, the maximum variation between the inner and outermost ions is $5\%$, whereas the the $51$-ion experiments it amounts to $8.5\%$. This effect leads to imperfect initial state preparation and to measurement errors when collective spin rotations are used to align the desired measurement basis with the one accessible in the subsequent fluorescence measurement. For the 51-ion experiments, we characterize the state preparation and measurement errors by computing the fidelity overlap between the ideal single-qubit state and the state estimated from experimental data at $t=0$. For $F(\rho_i, \rho_e) = \left(\mathrm{tr}\sqrt{\sqrt{\rho_i} \rho_e \sqrt{\rho_i}}\right)^2$, where $\rho_e$ is the experimentally measured state and $\rho_i$ is the ideal state, we find and average single-qubit fidelity of $F(\rho_i, \rho_e)=0.98$.

\onecolumngrid
\newpage
\begin{center}
\textbf{\large Supplemental Material: Bounded-Error Quantum Simulation via Hamiltonian and Lindbladian Learning}
\end{center}
\vspace{1cm}

\setcounter{section}{0}
\renewcommand{\thesection}{S\arabic{section}}
\setcounter{figure}{0}
\renewcommand{\thefigure}{S\arabic{figure}}

\section{Re-parametrization}\label{app:regularization}
Re-parametrization of the dynamical model is realized by modifying the Hamiltonian or Lindbladian ansatz, for instance, by adding or removing operator terms or by promoting dependencies among parameters. Such dependency constraints can be incorporated by introducing a \emph{regularization} term into the least-squares problem of Eq.~(..) in the main text. This term acts as \emph{penalty} for deviating from these dependencies. In Bayesian learning, regularization arises naturally through the inclusion of prior distributions, which penalize parameter values deviating from prior expectations~\cite{Bairey2019,Waqar2023}. Indeed, Hamiltonian learning has been formulated explicitly within the Bayesian formalism~\cite{Bairey2019}.

\subsection{Tikhonov regularization}
Here, we employ \emph{Tikhonov regularization} to re-parametrize the dynamical model. Specifically, a matrix $G^{\mathrm{T}}$ encodes dependencies among the parameters $c_i$, defining a reduced set of parameters $\vec{c} \mapsto G^{\mathrm{T}}\vec{c} = \vec{c}_G$. The matrix $G$ is of size $n\times\tilde{n}$, with $n\geq \tilde{n}$. One requires the columns of $G = (\vec{g}_1,\dots ,\vec{g}_{\tilde{n}})$ to be orthonormal, i.e., $\langle \vec{g}_i \vert \vec{g}_j \rangle = \delta_{ij}$. Then, $G^TG=\id_{\tilde{n}\times\tilde{n}}$, and $GG^T=\sum_{i=1}^{\tilde{n}} \ketbra{\vec{g}_i}$ is a projector onto the support of $G^T$, and therefore $G$ is an \emph{isometry}. The regularized least-squares problem then takes the form
\begin{equation}\label{eq:LSQ_Reg}
    \vec{c}^{\,\mathrm{rec}}(\beta)
    = \arg\min_{\vec{x}}
    \left[
        \|\tilde{M}\vec{x} - \tilde{\vec{b}}\|_2^2
        + \beta \|(\mathds{1} - GG^{\mathrm{T}})\vec{x}\|_2^2
    \right],
\end{equation}
where the second term penalizes deviations of the reconstructed parameters from the dependencies encoded in $G^T$. The regularization parameter $\beta$ controls the penalty. For $\beta=0$ one recovers the non-regularized solution, while in the limit $\beta\rightarrow\infty$ the dependencies encoded in $G^T$ are strictly enforced. For instance, in the analysis of the ten-qubit experimental data, we use $\vec{g}\propto (1/\abs{i-j}^{\alpha}: i=1,\dots,N; j=i+1,\dots,N)^T$ to promote solutions close to an algebraic decay. Then,
\begin{equation}
    G = \mqty(\vec{g} & \vec{0} \\ \vec{0}& \id).
\end{equation}
\subsection{Choosing the regularization parameter $\beta$}
A regularization that promotes inadequate dependencies between reconstructed parameters may introduce a bias. To distinguish adequate from inadequate dependencies we adopt the following strategy~\cite{Olsacher2025}: We consider the singular value decomposition of the concatenated matrix $[\vec{M},-\vec{b}]$. Its right-singular vector corresponding to the smallest singular value, $\lambda_1$, contains the learned Hamiltonian and Lindbladian parameters, while $\lambda_1$ can be interpreted as a learning error, similar to the residual norm $\norm{\vec{r}}_2$, discussed in the main text. The gap to the second-smallest singular value, $\lambda_2$, can be interpreted as the noise robustness of the reconstructed parameters. For large numbers of parameter this gap is typically small, and regularization increases this gap by reducing the number of effective parameters~\cite{Olsacher2025,Bairey2019}. Thus, when comparing different parametrization and tuning the specific values of $\beta$, one aims to maximize the gap, but not at the cost of significantly lifting the smallest singular value, i.e., of introducing a bias. This is achieved by minimizing the ratio $\lambda_1/\lambda_2$. In the analysis of the ten-qubit experimental data, the gap saturates at roughly $\beta=2.5\times 10^{-3}$, while $\lambda_1$ remains approximately constant, even for much larger values of $\beta$.

\section{Error bounds}\label{app:ErrorBounds}
\subsection{Correlated error models}
Consider the Hamiltonian model
\begin{equation}
    H'(\vec{g})=H+V(\vec{g})
\end{equation}
as in Eq.~(6) in the main text, where $\vec{g}\sim\mathcal{N}(0,\Sigma)$, with general covariance $\Sigma\geq 0$. We can transform to independent standard Gaussian variables thought the affine transformation $\vec{\eta}=\Sigma^{-\frac{1}{2}}\vec{g}$, with $\eta\sim\mathcal{N}(0,1)^M$. Then, 
\begin{equation}
V(\vec{g})=\sum_{i=1}^Mg_ih_i=\sum_{i,j=1}^M \Sigma^{\frac{1}{2}}_{ij} \eta_j h_i=   \sum_{j=1}^M \eta_j k_j, 
\end{equation}
where $k_j=\sum_{i=1}^M\Sigma^{\frac{1}{2}}_{ij} h_i$, and $\norm{k_j}\leq \sum_{i=1}^M\abs*{\Sigma^{\frac{1}{2}}_{ij}}$, as $\norm*{h_i}\leq 1$. Then, using the moment generating function of a multivariate Gaussian, the bound on the expected error in Eq.~(12) of Ref.~\cite{Cai2024} becomes
\begin{align}
    \abs{\expval{O_t}-\mathds{E}[\expval{O_t(\vec{g})}]}&\leq
    [e^{2t^2\sum_{j=1}^M (\sum_{i=1}^M\abs*{\Sigma^{\frac{1}{2}}_{ij}})^2}-1]\norm*{O}=[e^{2t^2\norm*{\Sigma^{\frac{1}{2}}}_{L_{1,2}}^2}-1]\norm*{O}.
\end{align}
A similar derivation as in Ref.~\cite{Cai2024} then yields the corresponding bounds on the concentration of the error around its expected value.

\subsection{Lieb-Robinson bounds}\label{sec:lieb-robinson}
Both our bounded error quantum simulation results and one of the Hamiltonian learning protocols we used (the derivative method) rely on the existence of approximate light-cones for the underlying dynamics. The standard tool to show the existence and bound the size of such lightcones are Lieb-Robinson bounds (LR-bounds)~\cite{Lieb1972}. In spite of the sizeable literature on the topic (see~\cite{AnthonyChen2023} for a recent review), our specific physical setup poses two challenges. First,  the system at hand could potentially have long-range interactions with a slow decay exponent, a regime that poses some challenges, as the so-called Lieb-Robinson velocity is then no longer necessarily independent of the system's size~\cite{Sweke2019}. To address this, we can resort to~\cite{Sweke2019}.

The second challenge posed by our particular setup is the presence of \emph{collective dephasing} in the system. More precisely, define the jump operator on $N$ qubits given by
\[
J_z \;=\; \tfrac12 \sum_{i=1}^{N} \sigma_z^{(i)}.
\]
The \emph{collective-dephasing Lindbladian} acts on any density matrix \(\rho\) as
\begin{equation}
\label{eq:Lindbladian}
\mathcal{L}_{\mathrm{deph}}(\rho)
\;=\;
\gamma
\Bigl(
    2 J_z \rho J_z
    - J_z^{2}\rho
    - \rho J_z^{2}
\Bigr),
\end{equation}
where \(\gamma\ge 0\) is the dephasing rate. The potential issue posed by the presence of this noise model is that the jump operator $J_z$ is global, i.e. it acts on all qubits at once. In contrast, LR-bounds typically require that all the elements of the generator of the evolution only act on a constant number of qubits, and it is natural to wonder if the noise could destroy finite light cones. We show that this is not the case and that, in fact, the presence of collective dephasing does not change the lightcones of the dynamics. To this end, we show LR bounds under the presence of collective dephasing by observing that such bounds follow from noting that  potentially unbounded single qubit terms do not change LR bounds.

\subsubsection{LR-bounds for systems with heavy tails and unbounded $1$-local terms}
We note that we slightly deviate from the notation in the main text in this subsection to make sure that our results follow a notation that is closer to that found in the LR-literature. Let us recall that our goal is to use LR-bounds to obtain bounded error simulation results and ensure that our Hamiltonian learning results apply. But we face two challenges: the fact that in our current physical setup we might have interactions with very long range and the presence of collective dephasing, which has nonlocal jumps. Thus, to address the first issue we discuss LR-bounds that give nontrivial bounds for any algebraic decay rate of the interactions, following the approach of~\cite{Sweke2019}. The price to pay will be that the LR-velocity is no longer a constant, but depends mildly on the system's size. In addition, to address collective dephasing, we will later see it suffices to show that LR-bounds are not influenced by (potentially unbounded, time-dependent) single qubit Hamiltonian terms in the generator. To show that, we follow the approach of~\cite{Nachtergaele2008}. Thus, in a nutshell, we follow and combine the approaches of~\cite{Sweke2019,Nachtergaele2008} to LR bounds for interactions with arbitrary tails and unbounded, time-dependent $1-$local terms.
We start by setting the stage for what is required to state and prove a LR bound:

\begin{enumerate}
\item\textbf{Lattice and metric.}
      A metric $d:\Lambda\times\Lambda\to\mathbb N_0$ on the set of qubits $\Lambda=\{1,\ldots,n\}$.

\item\textbf{Decay kernel.}
      A non-increasing $F:[0,\infty)\to(0,\infty)$ such that
      \begin{align}\label{equ:usual_decay}
      \|F\| :=\sup_{x\in\Lambda}\sum_{y\in\Lambda}F\bigl(d(x,y)\bigr)<\infty,
      \\
      C :=\sup_{x,y\in\Lambda}\sum_{z\in\Lambda}
           \frac{F\bigl(d(x,z)\bigr)F\bigl(d(y,z)\bigr)}
                {F\bigl(d(x,y)\bigr)}<\infty.
      \end{align}
      For any $\mu>0$ set $F_\mu(r)=e^{-\mu r}F(r)$, whence
      $\|F_\mu\|\le\|F\|$ and $C_\mu\le C$. A typical choice for $F$ on a $D$ dimensional lattice is $(1+d(x,y))^{-D-1}$
      Alternatively, in the case of long-range interactions we can follow~\cite{Sweke2019} and consider a kernel of the form $G_\eta(x)=(1+x)^{-\eta}$ for $\eta>0$ that satisfies instead for some constant $p_1>0$:
      \begin{align}\label{equ:slow_decay}
      \sup_{x,y\in\Lambda}\sum_{z\in\Lambda}
           \frac{G_\eta\bigl(d(x,z)\bigr)G_\eta\bigl(d(y,z)\bigr)}
                {G_\eta\bigl(d(x,y)\bigr)}\leq p_1/N_{\Lambda},
      \end{align}
      where 
      \begin{align}
      N_\Lambda=1/\sup_{x}\sup_{y\not= x}\sum(1+d(x,y))^{-\eta}
      \end{align}
We discuss LR bounds assuming both Eq.~\eqref{equ:slow_decay} and Eq.~\eqref{equ:usual_decay}. But note that, for a $D$ dimensional lattice we have that $N_\Lambda=\mathcal{O}(n^{\tfrac{\eta}{D}-1})$ for $\eta<D$~\cite{Kastner2011}.

\item\textbf{Time-dependent interaction.} For the sake of generality, we model both the Lindbladian and Hamiltonian part as time-dependent. For some subset $Z\subset \Lambda$ we let $\mathcal A_Z$ be the algebra of operators supported on $Z$. In addition, we have that $2^{\Lambda}$ is the set of all subsets of $\Lambda$.
      \begin{itemize}
      \item Hamiltonian part:
            for each subset of $[n]$ we assign an interaction
            $\Phi:\mathbb R\times 2^{\Lambda}\to\mathcal A_\Lambda$
            with $\Phi(t,Z)=\Phi(t,Z)^\dagger\in\mathcal A_Z$.
      \item Dissipative part:
            for each finite $Z\subset\Lambda$ we also assing a family of jump operators
            $\{L_a(t,Z)\}_{a=1}^{N(Z)}\subset\mathcal A_Z$.
      \end{itemize}
      For $A\in\mathcal A_Z$ set
      \begin{equation*}
        \Psi_Z(t)(A)=
          i\bigl[\Phi(t,Z),A\bigr]
          +\sum_{a=1}^{N(Z)}
             \Bigl(
               L_a^\dagger(t,Z)\,A\,L_a(t,Z)
               -\tfrac12\{L_a^\dagger(t,Z)L_a(t,Z),A\}
             \Bigr).
      \end{equation*}

\item\textbf{Truncated generator.}
      For $X\subset\Lambda$ define
      $\displaystyle
        \mathcal L_X(t)A:=\sum_{Z\subset X}\Psi_Z(t)(A)$. 

\item\textbf{Assumptions on regularity and decay.}
      \begin{enumerate}
      \item $t\mapsto\mathcal L_\Lambda(t)$ is norm-continuous for every finite $\Lambda$.
      \item Fast decay: define $\Psi'$ to be same generator as $\Psi$, except that we set $\Phi'(t,Z)=0$ for $|Z|=1$. I.e., we discard single body Hamiltonian terms. We define the following quantity for $\mu>0$:
\[
  \|\Psi\|_{t,\mu}:=\;
  \sup_{s\in[0,t]}
  \sup_{x,y\in\Lambda}
  \sum_{Z\ni x,y}
  \frac{\|\Psi_Z'(s)\|_{\mathrm{cb}}}
       {F_\mu\bigl(d(x,y)\bigr)}<\infty
\]
for all $t\geq 0$.
        \item Slow decay: define $\Psi'$ to be same generator as $\Psi$, except that we set $\Phi'(t,Z)=0$ for $|Z|=1$. I.e., we discard single body Hamiltonian terms. We define the following quantity for $\mu>0$:
        \[
              \|\Psi\|_{t,\eta}:=\;
              \sup_{s\in[0,t]}
              \sup_{x,y\in\Lambda}
              \sum_{Z\ni x,y,|Z|\not=1}
              \frac{\|\Psi_Z'(s)\|_{\mathrm{cb}}}
                   {G_\eta\bigl(d(x,y)\bigr)}<\infty
            \]
            for all $t\geq0$.
      \end{enumerate}

\end{enumerate}
We also need the following concepts:
\begin{enumerate}
    \item \emph{Completely bounded (cb) norm:}
    Let $\mathcal A$ and $\mathcal B$ be $C^{\ast}$–algebras and
    $T:\mathcal A\!\to\!\mathcal B$ a linear map.
    For every $n\in\mathbb N$ denote by
    $T^{(n)}:=T\otimes\mathrm{id}_{M_n}$ the \emph{$n$-fold amplification}
    acting on $\mathcal A\otimes M_n(\mathbb C)$.
    The \emph{completely bounded norm} of $T$ is
    \[
       \|T\|_{\mathrm{cb}}
       :=\sup_{n\ge 1}\,\bigl\|T^{(n)}\bigr\|\,.
    \]
    In finite dimensions (our qubit setting) this supremum is always attained
    and \(\|T\|_{\mathrm{cb}}\le D\|T\|\) with \(D=\dim\mathcal A\).
    \item \emph{Local maps that vanish on the identity:}
    For a finite subset \(X\subset\Lambda\) let
    \[
       \mathcal B_X
       :=\Bigl\{\,T\in\mathcal B(\mathcal A_X)\;
                 \Big|\,
                 T(1_{\!X})=0\Bigr\},
    \]
    the subspace of completely bounded maps on \(\mathcal A_X\) that \emph{annihilate the identity}. 
\end{enumerate}
Now that we have set the stage, let us prove our LR-bounds. It will essentially boil down to combining the proofs of~\cite{nachter1,Nachtergaele2008,Sweke2019}. The difference to~\cite{nachter1} is that we consider time-dependent Hamiltonian terms and Lindlbladians, but the same proof follows verbatim, as already remarked in~\cite{nachter1,Nachtergaele2008}.

Now let $\Lambda=\{1,\dots ,n\}$ be the finite qubit lattice endowed with a
metric $d$, let $F_\mu$ or $G_\eta$ be a decay kernel
satisfying~\eqref{equ:usual_decay} or~\eqref{equ:slow_decay}, and let the
time–dependent generator
\[
  \mathcal L_\Lambda(t)=\sum_{Z\subset\Lambda}\Psi_Z(t)
\]
be built from the Hamiltonian interactions $\Phi(t,Z)$ and jump
operators $L_a(t,Z)$.  Write
$\gamma_{t,s}^\Lambda=\exp\!\bigl(\int_{s}^{t}\! \mathcal
L_\Lambda(\tau)\,d\tau\bigr)$ for the resulting propagator.  Define the corresponding time-dependent semi\-norms
\begin{align*}
  \|\Psi\|_{t,\mu}':=
  \sup_{s\in[0,t]}\;
  \sup_{x,y\in\Lambda}
  \sum_{\substack{Z\ni x,y}}
  \frac{\|\Psi_Z'(s)\|_{\mathrm{cb}}}{F_\mu(d(x,y))},\\
  \|\Psi\|_{t,\eta}':=
  \sup_{s\in[0,t]}\!
  \sup_{x,y\in\Lambda}
  \sum_{\substack{Z\ni x,y}}
  \frac{\|\Psi_Z'(s)\|_{\mathrm{cb}}}{G_\eta(d(x,y))}.
\end{align*}

\noindent
\textbf{Fast-decay case.}
Assume~\eqref{equ:usual_decay} and $\|\Psi\|_{t,\mu}'<\infty$ for some
$\mu>0$.  
For any disjoint supports $X,Y\subset\Lambda$, any $K\in\mathcal B_X$
and $B\in\mathcal A_Y$,
\begin{equation}\label{eq:LR-fast}
  \bigl\|\,K\,\gamma_{t,s}^\Lambda(B)\bigr\|
  \;\le\;\frac{\|K\|_{\mathrm{cb}}\|B\|}{C_\mu}\,
  e^{\,C_\mu\|\Psi\|_{t,\mu}'(t-s)}
  \sum_{x\in X}\sum_{y\in Y} F_\mu\!\bigl(d(x,y)\bigr).
\end{equation}

\noindent
\textbf{Slow-decay case.}
Assume~\eqref{equ:slow_decay} and
$\|\Psi\|_{t,\eta}'<\infty$ for some $\eta>0$.  
Then for the same data $X,Y,K,B$
\begin{equation}\label{eq:LR-slow}
  \bigl\|\,K\,\gamma_{t,s}^\Lambda(B)\bigr\|
  \;\le\;p_1^{-1}\|K\|_{\mathrm{cb}}\|B\|\,
  e^{\,p_1\|\Psi\|_{t,\eta}'(t-s)/N_\Lambda}\sum_{x\in X}\sum_{y\in Y} G_\eta\!\bigl(d(x,y)\bigr).
\end{equation}
Consequently the Lieb–Robinson velocity is bounded by  
\(
  v_{\text{\tiny LR}}(\mu)\;=\;C_\mu\|\Psi\|_{t,\mu}'/\mu
\)
(respectively $v_{\text{\tiny LR}}(\eta)=p_1\|\Psi\|_{t,\eta}'/N_{\Gamma}$),
and \emph{depends only on multi-qubit couplings and jumps;
all single-qubit Hamiltonian terms have been projected out}.

\subsubsection{LR-bounds under collective dephasing}\label{subsub:dephasing}
We now show that collective dephasing does not change LR-bounds. To see this, note that collective dephasing can be viewed as a unitary evolution driven by a \emph{random, fluctuating, single qubit} Hamiltonian
\begin{align}\label{equ:Hamiltonian_dephasing}
V(t)\;=\;g(t)\,J_z,
\qquad
J_z=\tfrac12\sum_{i=1}^{N}\sigma_z^{(i)},
\end{align}
where the scalar noise \(g(t)\) is a real‐valued Wiener (Brownian)
increment with
\[
\mathbb{E}[g(t)]=0,
\qquad
\mathbb{E}[g(t)\,g(t')]=2\gamma\,\delta(t-t').
\]
Consider the (random) Hamiltonian $H'=H+g(t)V$. One can show that (see, e.g., Eq.~(9) in Ref.~\cite{Kiely2021} for a derivation) the expected time evolution
\begin{align}
\rho_t=\mathbb{E}_g(e^{itH'}\rho e^{-itH'})
\end{align}
satisfies the Lindblad equation:
\begin{align}
\frac{d}{dt}\rho_t=\mathcal{L}(\rho_t),\quad \mathcal{L}(\rho)=i[H,\rho]+\gamma\mathcal{L}_{\mathrm{deph}}(\rho).
\end{align}
Thus, we can see collective dephasing as arising as the average dynamics of a time-dependent random $1$-qubit Hamiltonian. However, as one would intuitively expect and as formalized above, single-qubit terms cannot affect the speed of propagation in a spin chain, no matter how large they are. Thus, as collective dephasing arises as the average over $1$-qubit Hamiltonian dynamics, it can also not increase the speed of propagation.
More formally, assume that for the local observable $O$ time-evolved by $H'$, $O_g(t)$, we have a LR-bound of the form
\begin{align}
\|[X,O_g(t)\|\leq \|O\|\|X\|\tau(r,t),
\end{align}
for some function $\tau(r,t)$ that decays in $r$ and independent of the time-dependent potential $g$, as it is just a $1$-qubit Hamiltonian. Then we claim that for $O'(t)$ the observable evolved by $\mathcal{L}'(O)=i[H,O]+\mathcal{L}_{\mathrm{deph}}(O)$ we have:
\begin{align}
    \|[X,O'(t)\|\leq \|O\|\|X\|\tau(r,t),
\end{align}
i.e., the same bound holds. This follows from a simple triangle inequality, as 
\begin{align}
     &\|[X,O'(t)\|=\|[X,\mathbb{E}_g(O_g(t))\|\leq \\ &\mathbb{E}_g\|[X,O_g(t)\|\leq \|O\|\|X\|\tau(r,t).
\end{align}
We formulated the bound for the commutator version of the LR bound, but the same argument extends to other versions as well. But, importantly, we see that collective dephasing does not hinder us from using LR bounds both for learning and the bounded error quantum simulation.

That being said, the effect of these two types of families of noise is qualitatively different. To see this, we can expand the generator of collective dephasing as:
\begin{align}
  \mathcal{L}_{\mathrm{deph}}(\rho)=\sum_{i,j=1}^n[\sigma_z^{(i)},[\sigma_z^{(j)},\rho]],
\end{align}
whereas single-qubit dephasing can be expanded as $\mathcal{L}_{\mathrm{deph},1}(\rho)=\sum_{i=1}^n[\sigma_z^{(i)},[\sigma_z^{(i)},\rho]]$. Thus, we see that the number of terms acting nontrivially on a region of size $k$ is $k^2$ for collective dephasing and $k$ for single qubit dephasing. We conclude that if both have comparable rates, then collective dephasing can generate a decay of expectation values of observables at a rate that is proportional to the square of the size of the support, whereas local dephasing only induces a rate that is linear in the size of the support. 

\section{Fluctuation bounds under stochastic Hamiltonians}

Let us define the random variable
\begin{equation}
    x_{\vec{g}}(t)=\expval{O'_{t}(\vec{g})}-\mathds{E}\!\left[\expval{O'_{t}(\vec{g})}\right]
\end{equation}
describing the distribution of $\expval{O'_{t}(\vec{g})}$ around its expected value. In the following, we will use a short-time expansion for $x_{g}(t)$ up to ${\cal O}\left(t^{2}\right)$ and derive high-probability bounds using (i) sub-Gaussian concentration for linear terms and (ii) the Hanson-Wright (HW) inequality for quadratic terms.

In Appendix~\ref{app:expansion} we show that $x_{\vec{g}}(t)$ can be expanded as
\begin{equation}\label{eq:expansion}
x_{\vec{g}}(t)=t \vec{b}_{t}^{\top} \vec{g}-\frac{t^{2}}{2}\left(\vec{g}^{\top} D(O) \vec{g}-\mathbb{E}[\vec{g}^{\top} D(O) \vec{g}]\right)+\mathcal{O}\left(t^{3}\right),
\end{equation}
where we have defined $\vec{b}_{t}=\vec{b}-\frac{t}{2} \vec{s}$, with entries
\begin{equation}
b_{i}=i \Tr(\rho_{0}\left[V_{i}, O\right]), \quad s_{j}=\Tr\left(\rho_{0}\left[\left[H, V_{j}\right], O\right]\right),
\end{equation}
and
\begin{equation}
    D_{i j}(O)=\Tr\left(\rho_{0}\left[V_{i},\left[V_{j}, O\right]\right]\right).
\end{equation}
The expansion coefficients above involve expectation values of few-body observables evaluated in the input state, $\rho_0$. Since $\rho_0$ is typically a product state, which can be reconstructed efficiently via state tomography, these expectation values and the resulting expansion coefficients can be efficiently measured. Next, we will derive high-probability bounds on $\abs{x_{\vec{g}}(t)}$ based on these expansion coefficients.

Let us first consider the linear term in Eq.~\eqref{eq:expansion}. If the $g_{i}$ are independent, and mean-zero Gaussian variables with variance $\delta^{2}$, then for any fixed vector $\vec{u}$ it holds that~\cite{Vershynin2018}
\begin{equation}
    \mathbb{P}(|\vec{u} \cdot \vec{g}|>x) \leq 2 \exp \left(-c \frac{x^{2}}{\delta^{2}\norm{\vec{u}}_{2}^{2}}\right).
\end{equation}
Thus, with $\vec{u}=\vec{b}_{t}$ we obtain that with probability at least $1-\eta / 2$
\begin{equation}\label{eq:linearBound}
\abs{\vec{b}_{t}\cdot\vec{g}} \leq \delta\norm{\vec{b}_t}_{2} \sqrt{\frac{1}{c} \log \frac{4}{\eta}}.
\end{equation}

To bound the quadratic term we will employ the Hanson-Wright inequality which states a general concentration inequality for quadratic forms of independent Gaussian random variables $g_i$ with variance $\delta^{2}$~\cite{Rudelson2013}
\begin{equation}
\mathbb{P}\left(\abs{\vec{g}^{\top} A \vec{g}-\mathbb{E}[\vec{g}^{\top} A \vec{g}] } > x \right) \leq 2 \exp \left(-\kappa \mu _{HW}\right)
\end{equation}
with
\begin{equation}
    \mu_{HW}=\min \left\{\frac{x^{2}}{\delta^{4}\|A\|_{\mathrm{HS}}^{2}}, \frac{x}{\delta^{2}\|A\|_{\mathrm{op}}}\right\},
\end{equation}
where $\norm{D}_{\rm HS}=\left(\sum_{i, j}\left|D_{i j}\right|^{2}\right)^{1 / 2}$ 
denotes the Hilbert-Schmidt norm, and $\norm{D}_{\rm op}=\lambda_{\rm max}(D)$ denotes the spectral norm, i.e., the largest singular value. Thus, for the quadratic term we obtain that with probability at least $1-\eta / 2$
\begin{equation}\label{eq:quadraticBound}
\abs{\vec{g}^{\top} D \vec{g}-\mathbb{E}[\vec{g}^{\top} D \vec{g}] } \leq \delta^{2} \max \left\{\|D\|_{\text {HS }} \sqrt{\frac{1}{\kappa} \log \frac{4}{\eta}}, \quad\|D\|_{\text {op }} \frac{1}{\kappa} \log \frac{4}{\eta}\right\}.
\end{equation}

From Eq.~\eqref{eq:expansion}, using the triangle inequality, and splitting the failure budget as $\eta / 2+\eta / 2$, both events in Eq.~\eqref{eq:linearBound} and Eq.~\eqref{eq:quadraticBound} hold simultaneously with probability at least $1-\eta$ (see Appendix~\ref{app:riskallocation} for comments on the allocation of risk level). Thus we obtain as a final bound
\begin{equation}\label{eq:HW}
\bigl|\mathds{E}\!\left[\expval{O'_{t}(\vec{g})}\right]-\expval{O'_{t} (\vec{g})}\bigr| 
\leq \delta t \norm{b_t}_2\sqrt{\frac{1}{c}\log{\frac{4}{\eta}}} + \frac{t^2}{2}\delta^2\max\qty{\norm{D}_{\rm HS}\sqrt{\frac{1}{\kappa}\log{\frac{4}{\eta}}}, \norm{D}_{\rm op}\frac{1}{\kappa}\log{\frac{4}{\eta}}},
\end{equation}
where $\kappa \sim O(1)$ is a universal constant whose precise numerical values varies in the literature and only affect the tightness of the exponential tail. In practice, $\kappa$ can be slightly calibrated to obtain sharper fits to empirical data. Let us also emphasize that this bound also holds in case $g_i$ are sub-Gaussian with variance proxy $\delta^2$.

\subsection{Short-time evolution}\label{app:expansion}
Using the Baker-Campbell-Hausdorff formula we can expand the expectation value as
\begin{equation}
    \expval{O'_{t}(\vec{g})}=\Tr\left[\rho_{0} O\right] -i t \Tr\left[\rho_{0}\left[O, H'\right]\right] -\frac{t^{2}}{2} \Tr\left[\rho_{0}\left[H',\left[H', O\right]\right]\right]+\mathcal{O}\left(t^{3}\right).
\end{equation}
Using $H'=H+\sum_{j} g_{j} V_{j}$ and the linearity of the commutator, we can expand this as
\begin{align}
\expval{O'_{t}(\vec{g})}= & \Tr\left(\rho_{0} O\right) \notag -i t \Tr\left(\rho_{0}[O, H]\right)-i t \sum_{j} g_{j} \Tr\left(\rho_{0}\left[O, V_{j}\right]\right) \notag -\frac{t^{2}}{2} \Tr\left(\rho_{0}[H,[H, O]]\right) \notag\\
&-\frac{t^{2}}{2} \sum_{j} g_{j} \Tr\left(\rho_{0}\left[H,\left[V_{j}, O\right]\right]+\rho_{0}\left[V_{j},[H, O]\right]\right) -\frac{t^{2}}{2} \sum_{j, k} g_{j} g_{k} \Tr\left(\rho_{0}\left[V_{j},\left[V_{k}, O\right]\right]\right)+\mathcal{O}\left(t^{3}\right).
\end{align}
Defining the real coefficients
\begin{equation}
b_{j} =i \Tr\left(\rho_{0}\left[V_{j}, O\right]\right), \qquad s_{j} =\Tr\left(\rho_{0}\left[\left[H, V_{j}\right], O\right]\right), \qquad D_{j k} =\Tr\left(\rho_{0}\left[V_{j},\left[V_{k}, O\right]\right]\right),
\end{equation}
we finally get
\begin{equation}
\expval{O'_{t}(\vec{g})}= \Tr\left(\rho_{0} O\right)+i t \Tr\left(\rho_{0}[H, O]\right)+t \sum_{j} g_{j} b_{j} -\frac{t^{2}}{2} \Tr\left(\rho_{0}[H,[H, O]]\right)-\frac{t^{2}}{2} \sum_{j} g_{j} s_{j}-\frac{t^{2}}{2} g^{\top} D g+\mathcal{O}\left(t^{3}\right).
\end{equation}
Next, we consider the average over disorder. Since $\mathbb{E}\left[g_{j}\right]=0$ and $\mathbb{E}\left[g_{j} g_{k}\right]=\delta^{2} \delta_{j k}$, we have
\begin{equation}
\mathbb{E}\left[\expval{O'_{t}(\vec{g})}\right]= \Tr\left(\rho_{0} O\right)+i t \Tr\left(\rho_{0}[H, O]\right)-\frac{t^{2}}{2} \Tr\left(\rho_{0}[H,[H, O]]\right) -\frac{t^{2}}{2} \delta^{2} \Tr(D)+\mathcal{O}\left(t^{3}\right)
\end{equation}
Putting everything together, we finally obtain Eq.~\eqref{eq:expansion}
\begin{equation}
x_{\vec{g}}(t)=t \vec{b}_{t}^\top \vec{g}-\frac{t^{2}}{2}\left(\vec{g}^{\top} D \vec{g}-\mathbb{E}\left[\vec{g}^{\top} D \vec{g}\right]\right)+\mathcal{O}\left(t^{3}\right),    
\end{equation}
where $\vec{b}_{t}:=\vec{b}-\frac{t}{2} \vec{s}$.

\subsection{Union bound and allocation of the risk level}\label{app:riskallocation}
We need both the linear and quadratic fluctuation bounds to hold simultaneously. Let
\begin{align}
    E_{1}&\equiv\left\{\left|g \cdot b_{t}\right| \leq \text { linear cutoff }\right\}, \\
    E_{2}&\equiv\left\{\left|\Delta_{\text {quad }}\right| \leq \text { quadratic cutoff }\right\}
\end{align}
with $\Delta_{\text {quad }}=\vec{g}^{\top} D \vec{g}-\mathbb{E}\left[\vec{g}^{\top} D \vec{g}\right]$. The complements $E_{1}^{c}, E_{2}^{c}$ denote violations of the respective bounds. With the De Morgan's law, $\left(E_{1} \cap E_{2}\right)^{c}=E_{1}^{c} \cup E_{2}^{c}$ and the Boole's inequality, we get
\begin{align}
\operatorname{Pr}\left(E_{1} \cap E_{2}\right)=1-\operatorname{Pr}\left(E_{1}^{c} \cup E_{2}^{c}\right) \geq 1-\operatorname{Pr}\left(E_{1}^{c}\right)-\operatorname{Pr}\left(E_{2}^{c}\right)
\end{align}
{\it Splitting the risk level.---} By fixing the total risk level $\eta \in(0,1)$ and with $\operatorname{Pr}\left(E_{1}^{c}\right) \leq \eta_{1}$ and $\operatorname{Pr}\left(E_{2}^{c}\right) \leq \eta_{2}$, with $\eta_{1}+\eta_{2}=\eta$, we have
\begin{align}
\operatorname{Pr}\left(E_{1} \cap E_{2}\right) \geq 1-\eta
\end{align}
The most common choice is the symmetric split $\eta_{1}=\eta_{2}=\eta / 2$, yielding the $\log (4 / \eta)$ factors in the confidence bounds. More generally, the split can be asymmetric: e.g., if the linear term is dominant, one may assign a larger budget $\eta_{1}$ to it and a smaller $\eta_{2}$ to the quadratic part, slightly tightening the overall envelope. The rule of thumb criterion is to balance the contributions so that neither bound is unnecessarily loose given the problem parameters. Example of an uneven risk. Let's choose a total risk level $\eta \in(0,1)$ and an arbitrary split $\eta_{1}, \eta_{2}>0$ with $\eta_{1}+\eta_{2}=\eta$. The HW bound becomes
\begin{equation}
\left|x_{\vec{g}}(t)\right|  \leq \delta t\left\|b_{t}\right\|_{2} \sqrt{\frac{1}{c} \log \frac{2}{\eta_{1}}} +\frac{t^{2}}{2} \delta^{2} \max \left\{\|D\|_{\text {HS }} \sqrt{\frac{1}{\kappa} \log \frac{2}{\eta_{2}}},\|D\|_{\text {op }} \frac{1}{\kappa} \log \frac{2}{\eta_{2}}\right\}+\mathcal{O}\left(t^{3}\right).
\end{equation}
{\it Risk split.---} Let the total failure budget be $\eta \in(0,1)$ and write $\eta_{2}=\eta-\eta_{1}$ with $\eta_{1} \in(0, \eta)$. The union-bound envelope reads
$$
\left|x_{\vec{g}}(t)\right| \leq L\left(\eta_{1}\right)+Q\left(\eta-\eta_{1}\right)+\mathcal{O}\left(t^{3}\right)
$$
with
$$
L\left(\eta_{1}\right)=\delta t\left\|b_{t}\right\|_{2} \sqrt{\frac{1}{c} \log \frac{2}{\eta_{1}}}, \quad Q\left(\eta_{2}\right)=\frac{t^{2}}{2} \delta^{2} \max \left\{\|D\|_{\mathrm{HS}} \sqrt{\frac{1}{\kappa} \log \frac{2}{\eta_{2}}}, \quad\|D\|_{\mathrm{op}} \frac{1}{\kappa} \log \frac{2}{\eta_{2}}\right\} .
$$
Define constants
$$
A:=\frac{\delta t}{\sqrt{c}}\left\|b_{t}\right\|_{2}, \quad B_{\mathrm{HS}}:=\frac{t^{2}}{2} \frac{\delta^{2}}{\sqrt{\kappa}}\|D\|_{\mathrm{HS}}, \quad B_{\mathrm{op}}:=\frac{t^{2}}{2} \frac{\delta^{2}}{\kappa}\|D\|_{\mathrm{op}} .
$$
Then the we have to find the minimum of
$$
\mathcal{B}\left(\eta_{1}\right)=A \sqrt{\log \frac{2}{\eta_{1}}}+\max \left\{B_{\mathrm{HS}} \sqrt{\log \frac{2}{\eta-\eta_{1}}}, \quad B_{\mathrm{op}} \log \frac{2}{\eta-\eta_{1}}\right\}, \quad \eta_{1} \in(0, \eta).
$$
{\it Switch point between branches.---} Let $\ell\left(\eta_{2}\right):=\log \frac{2}{\eta_{2}}$. The HS and operator branches are equal when
$$
B_{\mathrm{HS}} \sqrt{\ell\left(\eta_{2}\right)}=B_{\mathrm{op}} \ell\left(\eta_{2}\right) \quad \Longleftrightarrow \quad \ell\left(\eta_{2}\right)=\left(\frac{B_{\mathrm{HS}}}{B_{\mathrm{op}}}\right)^{2} .
$$
If $\eta_{2}^{\star}:=2 \exp \left(-\frac{B_{\mathrm{HS}}^{2}}{B_{\mathrm{op}}^{2}}\right)$ lies in $(0, \eta)$, the switch occurs at $\eta_{1}^{\mathrm{sw}}=\eta-\eta_{2}^{\star}$; otherwise one branch dominates on the whole interval.\\
{\it Stationary conditions within each branch.---} If the minimizer lies in the HS branch,
$$
\mathcal{B}^{\prime}\left(\eta_{1}\right)=-\frac{A}{2 \eta_{1} \sqrt{\log \left(2 / \eta_{1}\right)}}+\frac{B_{\mathrm{HS}}}{2\left(\eta-\eta_{1}\right) \sqrt{\log \left(2 /\left(\eta-\eta_{1}\right)\right)}}=0
$$
If it lies in the operator branch,
$$
\mathcal{B}^{\prime}\left(\eta_{1}\right)=-\frac{A}{2 \eta_{1} \sqrt{\log \left(2 / \eta_{1}\right)}}+\frac{B_{\mathrm{op}}}{\eta-\eta_{1}}=0
$$
Each equation admits a unique solution in its interior domain (if it exists) and can be solved numerically.

\section{Fidelity bounds}
Let $\ket{\psi_0}$ be a fixed input state and define the ideal and noisy
time–evolved states
\begin{equation}
\ket{\psi(t)} \;=\; e^{-iHt}\ket{\psi_0}, 
\qquad
\ket{\psi'(t,\vec g)} \;=\; e^{-iH'(\vec g)t}\ket{\psi_0}.
\end{equation}
The (state) fidelity is
\begin{equation}
F(t,\vec g) \;=\; 
\bigl|\braket{\psi(t)}{\psi'(t,\vec g)}\bigr|^2.
\end{equation}

For small noise–time product $\delta t \ll 1$, a Baker–Campbell–Hausdorff
expansion of $e^{iHt}e^{-iH'(\vec g)t}$ around the identity yields a
second–order expansion of $F(t,\vec g)$ of the form
\begin{equation}
1 - F(t,\vec g) 
    \;=\; t^2\,\vec g^{\top} C \,\vec g + \mathcal O(t^3),
\label{eq:fidelity-quadratic}
\end{equation}
where the real symmetric matrix $C\in\mathbb R^{M\times M}$ is
\begin{equation}
C_{ij} \;=\; \Re\bigl(\bra{\psi_0} V_i V_j \ket{\psi_0}\bigr).
\label{eq:C-def}
\end{equation}
In particular,
\begin{equation}
\mathbb E[1-F(t,\vec g)]
    \;=\; t^2\,\mathbb E[\vec g^{\top} C \vec g] 
    \;+\; \mathcal O(t^3)
    \;=\; t^2 \delta^2 \,\mathrm{Tr}(C) + \mathcal O(t^3),
\end{equation}
where we used $\mathbb E[g_i g_j] = \delta^2 \delta_{ij}$.

It is convenient to isolate the (centered) quadratic fluctuation
\begin{equation}
\Delta \;\equiv\;
\vec g^{\top} C \vec g - \mathbb E[\vec g^{\top} C \vec g],
\end{equation}
so that
\begin{equation}
1-F(t,\vec g)
    \;=\; t^2 \mathbb E[\vec g^{\top} C \vec g]
        + t^2 \Delta
        + \mathcal O(t^3),
\end{equation}
and hence
\begin{equation}
\bigl|(1-F(t,\vec g)) - \mathbb E[1-F(t,\vec g)]\bigr|
    \;=\; t^2|\Delta| + \mathcal O(t^3).
\label{eq:fidelity-fluctuation}
\end{equation}
Thus, any concentration inequality for the quadratic form $\vec g^{\top} C \vec g$
immediately yields a high–probability bound on the fidelity.

Let $C$ be real symmetric and let $\|\cdot\|_{\mathrm{HS}}$ and
$\|\cdot\|_{\mathrm{op}}$ denote the Hilbert–Schmidt and operator norms,
respectively.  For independent sub–Gaussian $g_i$ with variance proxy
$\delta^2$, the Hanson–Wright inequality implies that there exists a
universal constant $\kappa>0$ such that for all $u>0$
\begin{equation}
\mathbb P\!\left(
  \bigl|\Delta\bigr| > u
\right)
\;\leq\;
2\exp\!\left(
  -\kappa \min\left\{
    \frac{u^2}{\delta^4\|C\|_{\mathrm{HS}}^2},
    \frac{u}{\delta^2\|C\|_{\mathrm{op}}}
  \right\}
\right).
\label{eq:HW-Delta}
\end{equation}

Fix a failure probability $\eta\in(0,1)$ and choose $u$ such that the
right-hand side of \eqref{eq:HW-Delta} is at most $\eta$.  A convenient
choice is obtained by imposing
\begin{equation}
\kappa \min\left\{
\frac{u^2}{\delta^4\|C\|_{\mathrm{HS}}^2},
\frac{u}{\delta^2\|C\|_{\mathrm{op}}}
\right\}
\;\geq\; \log\frac{2}{\eta}.
\end{equation}
This leads to the two regimes
\begin{equation}
u_{\mathrm{HW}}(\eta) = 
\delta^2
\max\!\left\{
\|C\|_{\mathrm{HS}}
\sqrt{\frac{1}{\kappa}\log\frac{2}{\eta}},
\;
\|C\|_{\mathrm{op}}
\frac{1}{\kappa}\log\frac{2}{\eta}
\right\},
\label{eq:u-HW}
\end{equation}
for which
\begin{equation}
\mathbb P\!\left(
|\Delta| > u_{\mathrm{HW}}(\eta)
\right)
\;\leq\; \eta.
\end{equation}

Combining \eqref{eq:fidelity-fluctuation} and \eqref{eq:u-HW} we obtain,
with probability at least $1-\eta$,
\begin{equation}
\bigl|1 - F(t,\vec g) - \mathbb E[1-F(t,\vec g)]\bigr|
\leq t^2 u_{\mathrm{HW}}(\eta) + \mathcal O(t^3) = t^2 \delta^2
\max\!\left\{
\|C\|_{\mathrm{HS}}
\sqrt{\frac{1}{\kappa}\log\frac{2}{\eta}},
\;
\|C\|_{\mathrm{op}}
\frac{1}{\kappa}\log\frac{2}{\eta}
\right\}+ \mathcal O(t^3).
\end{equation}

Using $\mathbb E[1-F(t,\vec g)] = \mathcal O(M\delta^2 t^2)$, where
$M$ is the number of noisy terms, we obtain the following lower bound
on the fidelity, valid for $\delta t\ll 1$:
\begin{equation}
F(t,\vec g)
\;\geq\;
1
- t^2 \delta^2
\max\!\left\{
\|C\|_{\mathrm{HS}}
\sqrt{\frac{1}{\kappa}\log\frac{2}{\eta}},
\;
\|C\|_{\mathrm{op}}
\frac{1}{\kappa}\log\frac{2}{\eta}
\right\}  
- \mathcal O(M\delta^2 t^2),
\label{eq:fidelity-HW-linear}
\end{equation}
with probability at least $1-\eta$.

Equivalently, for small $t$ one may rewrite the bound in exponential
form as
\begin{equation}
F(t,\vec g)
\;\geq\;
\exp\!\left[
-\,t^2 \delta^2
\max\!\left\{
\|C\|_{\mathrm{HS}}
\sqrt{\frac{1}{\kappa}\log\frac{2}{\eta}},
\;
\|C\|_{\mathrm{op}}
\frac{1}{\kappa}\log\frac{2}{\eta}
\right\} 
- \mathcal O(M\delta^2 t^2)
\right],
\label{eq:fidelity-HW-exp}
\end{equation}

where the $\mathcal O(M\delta^2 t^2)$ term collects the (typically
small) bias contribution $\mathbb E[1-F(t,\vec g)]$ at short times.  


\begin{thebibliography}{92}
\bibitem [{fig()}]{figNote}%
  \BibitemOpen
  \href@noop {} {}\bibinfo {note} {Note, that this is an eigenstate of the
  ideal model in Eq.~\eqref{eq:XYmodel}, but not an exact eigenstate of the
  learned Hamiltonian.}\BibitemShut {Stop}%
\end{thebibliography}
\end{document}